\documentclass[%
 reprint,
 superscriptaddress,
 amsmath,amssymb,
 aps,
]{revtex4-2}

\usepackage{graphicx}
\usepackage{dcolumn}
\usepackage{bm}
\usepackage{braket}
\usepackage{xcolor}
\usepackage{amsmath}
\usepackage{upgreek}
\usepackage{multirow} 
\usepackage{textalpha}
\usepackage{xr}  

\begin{document}

\preprint{APS/123-QED}

\title{Nuclear spin relaxation in zero- to ultralow-field magnetic resonance spectroscopy}%

\author{Florin Teleanu}
\affiliation{Department of Chemistry, New York University, New York, NY 10003, United States}
\affiliation{ELI-NP, “Horia Hulubei” National Institute for Physics and Nuclear Engineering, 30 Reactorului Street, Bucharest-Magurele, 077125, Ilfov, Romania}
\author{Anne M. Fabricant}
\altaffiliation{Current address: Department of Biosignals, Physikalisch-Technische Bundesanstalt (PTB), 10587 Berlin, Germany}
\affiliation{Institute of Physics, Johannes Gutenberg University of Mainz}
\affiliation{Helmholtz Institute Mainz, 55099 Mainz, Germany}
\affiliation{GSI Helmholtzzentrum für Schwerionenforschung, 64291 Darmstadt, Germany}
\author{Chengtong Zhang}
\affiliation{Department of Chemistry, New York University, New York, NY 10003, United States}
\author{Gary P. Centers}
\affiliation{Institute of Physics, Johannes Gutenberg University of Mainz}
\affiliation{Helmholtz Institute Mainz, 55099 Mainz, Germany}
\affiliation{GSI Helmholtzzentrum für Schwerionenforschung, 64291 Darmstadt, Germany}
\author{Dmitry Budker}
\affiliation{Institute of Physics, Johannes Gutenberg University of Mainz}
\affiliation{Helmholtz Institute Mainz, 55099 Mainz, Germany}
\affiliation{GSI Helmholtzzentrum für Schwerionenforschung, 64291 Darmstadt, Germany}
\affiliation{Department of Physics, University of California, Berkeley, CA 94720, USA}
\author{Danila A. Barskiy}
\altaffiliation{Current address: Frost Institute for Chemistry and Molecular Science, 
Department of Chemistry, University of Miami, Coral Gables, FL 33146, USA}
\affiliation{Institute of Physics, Johannes Gutenberg University of Mainz}
\affiliation{Helmholtz Institute Mainz, 55099 Mainz, Germany}
\affiliation{GSI Helmholtzzentrum für Schwerionenforschung, 64291 Darmstadt, Germany}
\author{Alexej Jerschow}
\email{Contact author: alexej.jerschow@nyu.edu}
\affiliation{Department of Chemistry, New York University, New York, NY 10003, United States}

\date{\today}

\begin{abstract}
Nuclear-magnetic-resonance experiments can interrogate a broad spectrum of molecular-tumbling regimes and can accurately measure interatomic distances in solution with sub-nanometer resolution. In the zero- to ultralow-field (ZULF) regime, population and coherence decay reveal nontrivial behavior due to strong coupling between nuclear spins. We note, in particular, the surprising effects that different resonances show different relaxation rates, depending on  (i) the (pre)polarizing magnet field and the shuttling trajectory to the detection region at nano- and microtesla fields, (ii) the strength of the measurement field, (iii) the detection method (single-channel or quadrature), and even (iv) the nutation angle induced by the excitation pulse. We describe herein experimental data of relaxation rates measured for a $^{13}$C-labeled formic acid sample, with an atomic-magnetometer-based ZULF setup, and develop a theoretical framework to explain the detected effects and extract molecular properties. The observed effects could be used for spectral assignment, for the establishment of specific motional regimes, for image contrast, and for the characterization of relaxation processes at nano- to microtesla magnetic fields.   

\end{abstract}

\maketitle


In nuclear magnetic resonance (NMR) experiments of liquids, molecular tumbling produces fluctuating interactions, which induce nuclear spin flips and ultimately establish equilibrium spin populations\cite{abragam_principles_1961,levitt_spin_2013,kuprov_spin_2023,kowalewski_nuclear_2019}. 
Monitoring these recovery processes, described by measured relaxation rates, provides a characterization of sub-millisecond molecular dynamics\cite{charlier_protein_2016,stenstrom_how_2022}, inter-particle distances via the Overhauser effect\cite{overhauser_polarization_1953,noggle_nuclear_1971,kumar_two-dimensional_1980}, relative orientations and angles through cross-correlated interactions\cite{kumar_cross-correlations_2000}, and the local electric and magnetic fields impacting quadrupolar nuclei\cite{jerschow_nuclear_2005}.

The relaxometry framework is typically used in the context of high-field NMR, since chemical shift resolution and sensitivity improve significantly with the strength of the external field\cite{cavanagh_protein_1995,abragam_principles_1961}. Field-cycling (FC) experiments\cite{kimmich_field-cycling_2018,bolik-coulon_comprehensive_2023} can be used to monitor relaxation rates (usually $T_1$) at much lower fields (down to hundreds of nT) using mechanical shuttling to move the sample between the field at which relaxation processes are allowed to progress, and the field at which signals are detected. 

Here, we analyze the relaxation rates measured at zero-to-ultralow fields (ZULF), where scalar coupling dominates over the Zeeman interaction\cite{barskiy_zero-_2025}, and build a theoretical framework for their interpretation and straightforward connection to internuclear distances and molecular tumbling. In contrast to high-field FC experiments, our experiments are performed by shuttling a prepolarized sample through a guiding field to the detection region for storage and subsequent measurement. The detection of nuclear spin signals is performed using optical atomic magnetometers\cite{budker_optical_2007,fabricant_proton_2024} and takes place at nT--$\upmu$T fields where Zeeman and scalar coupling interactions have similar strength. We observe several surprising and counter-intuitive effects in this regime, including the dependence of peak lifetimes on both pulse flip angles and on spin-system preparation by magnetic-field cycling, as well as the observation of different rates for different transitions in the ZULF spectra. Such behavior is distinctly different from that observed in high-field FC experiments\cite{bodenstedt_fast-field-cycling_2021,alcicek_zero-_2023,zhukov_field-cycling_2018,zhukov_assessment_2019}, and therefore represents a missing piece in regimes where inductive coils are inefficient for detection. We also provide a model-based description of polarization lifetimes in experiments performed on $^{13}$C-labeled formic acid at different ultralow fields and show how to derive molecular properties from observed effects.

\begin{figure}[htb]
\includegraphics[page=1, trim=30pt 25pt 50pt 50pt, clip, width=0.5\textwidth]{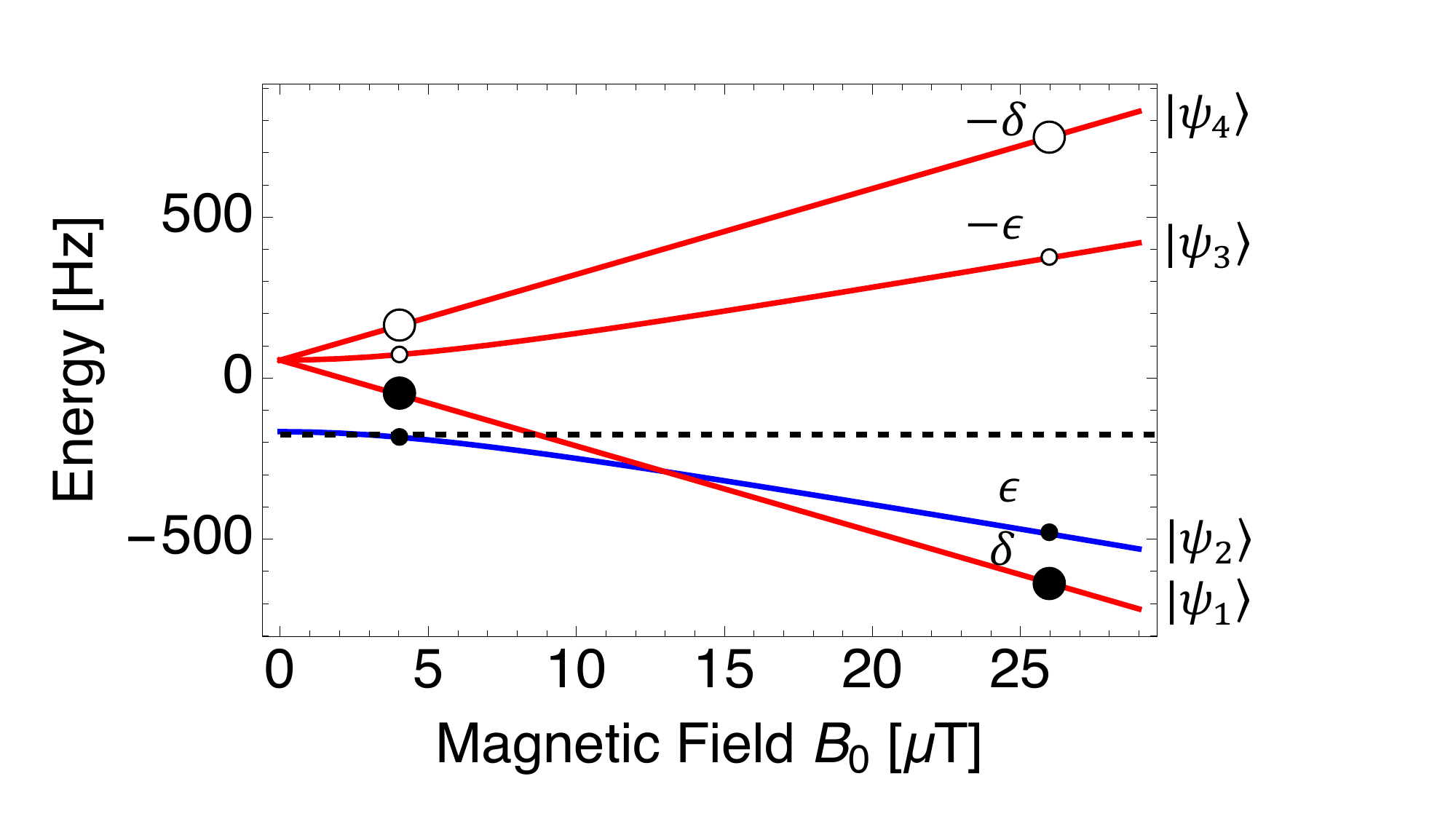}\caption{Breit-Rabi diagram showing the shifts of energy levels as a function of magnetic field. The populations have been adiabatically transferred from high to low field. Filled and empty circles represent overpopulated and underpopulated states, respectively. The diagram was calculated for  the [$^{13}$C-$^{1}$H] pair in formic acid ($J_{IS} = 222$\,Hz).}
\label{Breit_Rabi}
\end{figure}

Consider a heteronuclear two-spin-1/2 system with spin operator labels \textit{\{I, S\}} characterized by the $\gamma_{I}$ and $\gamma_{S}$ gyromagnetic ratios (in $\text{rad} \cdot \text{s}^{-1} \cdot $T$^{-1}$) and a scalar coupling constant $J_{IS}$ (in Hz). The active space can be described in terms of the field-dependent eigenstates $\{\ket{\psi_{i}}_{i=1,4}\}$ (See Appendix \ref{appenx:eigenstates}) which range from the high-field Zeeman basis to the singlet-triplet basis at zero field \cite{levitt_spin_2013}. In ZULF experiments, it is common to shuttle the sample from a (pre)polarizing region characterized by a higher homogeneous field $B_{p}$ to a region shielded from external sources of magnetic fields. Here, the main field $B_0$ is generated with a guiding solenoid that is controlled by the user. At the beginning of the experiment, the populations of the Zeeman states are:
\begin{equation}
    p_{\alpha\alpha}=\frac{1}{4}+\delta;p_{\alpha\beta}=\frac{1}{4}+\epsilon ; p_{\beta\alpha}=\frac{1}{4}-\epsilon; p_{\beta\beta}=\frac{1}{4}-\delta \,,
    \label{eq:populations}
\end{equation}
where $\delta=\frac{B_{p}(\gamma_{I}+\gamma_{S})}{8k_{B}T}$ and $\epsilon=\frac{B_{p}(\gamma_{I}-\gamma_{S})}{8k_{B}T}$ with $k_B$ as the Boltzmann constant and $T$ the temperature. The thermal equilibrium density operator at this field is given by $\hat{\rho}_{\text{eq}} (B_{p})  \approx \frac{1}{4k_{B}T}( - \gamma_{I}B_{p}\hat{I}_{z}-\gamma_{S}B_{p}\hat{S}_{z})$, where we neglect the identity matrix as it does not contribute to the observed signal. During shuttling, the high-field populations can be transferred adiabatically or suddenly to the eigenstates at the final field $B_0$. Based on our analyses (see Sections I and II in Supporting Information), we estimate that our experimental shuttling setup transfers populations adiabatically (Fig.\,\ref{Breit_Rabi}), such that the high-field density operator rewritten as $\hat{\rho}_{\text{eq}}{} =(\delta+\epsilon)\hat{I}_{z} +(\delta-\epsilon)\hat{S}_{z}$ transforms to

\begin{equation}
    \hat{\rho}_{\text{ad}} = \delta \hat{D}_{z} - \epsilon\left(\sin{(2\theta)}\hat{Z}_{x}+\cos{(2\theta)}\hat{Z}_{z}\right ) \,.
    \label{eq:rho_adiabatic}
\end{equation}
Here, $\theta = \frac{1}{2}\arctan\left (\frac{2\pi J_{IS}}{|\gamma_{I}-\gamma_{S}|B_{0}}\right )$ is known as the `mixing angle'\cite{pileio_singlet_2017} and we adopt the notation\cite{harris_zero-_2016} $\hat{D}_z = \hat{I}_z + \hat{S}_z$, $\hat{Z}_x = 2\hat{I}_x\hat{S}_x + 2\hat{I}_y\hat{S}_y$, $\hat{Z}_y = 2\hat{I}_x\hat{S}_y - 2\hat{I}_y\hat{S}_x$, $\hat{Z}_z = \hat{I}_z - \hat{S}_z$.

\begin{figure*}[htb]
\includegraphics[trim=0pt 85pt 0pt 80pt, clip, width=\textwidth]{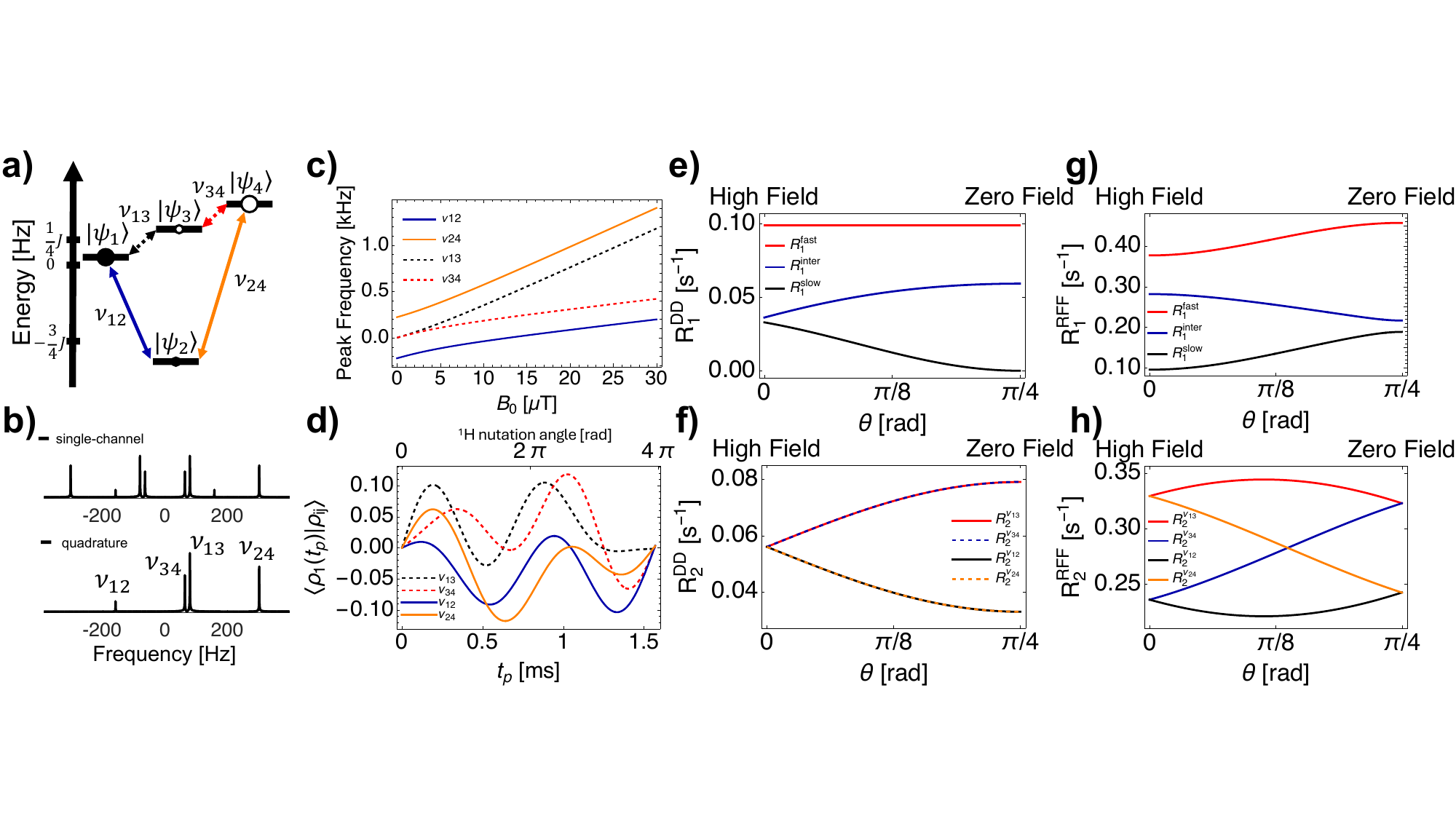}\caption{a) Energy-level diagram for the [$^{13}$C-$^{1}$H] system of $^{13}$C-labeled formic acid at ultralow field ($B_{0}=2.7\,\upmu$T and $J_{CH}=222$\,Hz). Observed transitions $\nu_{ij}$ connect nuclear-spin eigenstates, satisfying the selection rule $\Delta m_{f} = \pm 1$. b) Example of the simulated ULF spectrum corresponding to either single-channel detection or dual-channel quadrature detection. Reduced spectral crowding, absolute determination of peak frequencies, and increased signal amplitude are achieved using quadrature detection
c) Field dependence of the observable peak frequencies. d) Simulated nutation profile (Eqs.\,\ref{eq:nut13}--\ref{eq:nut24}) of the observable coherences $\hat{\rho}_{ij}=\ket{\psi_{i}}\bra{\psi_{j}}$ under a DC pulse along the $y$-axis, characterized by a magnetic field $B_{y}=30\,\upmu$T and total duration $t_p=1.6$\,ms; Theoretical relaxation rates of populations (e) and observable coherences (f) under dipolar interactions (DD) as a function of mixing angle $\theta$ for a [$^{13}$C-$^{1}$H] system in the fast-tumbling regime. The internuclear distance is $r_{\text{CH}}=1.1$\,\AA \ and the rotational correlation time is $\tau_{c}^{\text{DD}}=3.24$\,ps. Theoretical relaxation rates of populations (g) and observable coherences (h) under random field fluctuations (RFF), as a function of the mixing angle $\theta$ for a [$^{13}$C-$^{1}$H] system in the fast-tumbling regime. Parameters used for the simulations are $\omega_{\text{rms}}^{\text{C}}=85.32\cdot 10^3\,$rad$;\omega_{\text{rms}}^{\text{H}}=49.63 \cdot10^{3}\,$rad$; \kappa_{\text{CH}}=0.49$, and $\tau_{c}^{\text{RFF}}=19.34$\,ps.}
\label{nutation_R1_R2_DD_RFF}
\end{figure*}


Figure \ref{nutation_R1_R2_DD_RFF}a shows the energy-level diagram (along with population) of the {$^{13}$C-$^1$H} spin system, highlighting the observed transitions generated by superpositions of states satisfying the selection rule $\Delta m_f = \pm 1 $, with $m_f$ being the projection of the total spin angular momentum $F$ along the quantization axis\cite{stern_simulation_2023}. 
In order to convert the stationary density operator from Eq.\,\ref{eq:rho_adiabatic} into detectable magnetization, we apply short microsecond DC pulses along axes perpendicular to that of the small background magnetic field $B_{0}$ and measure the precession of polarization in the transverse plane. 
Depending on the experimental setup, detection can be performed with one or multiple magnetometers in configurations that allow for enhanced signal-to-noise ratios and unequivocal determination of the direction of precession with respect to the laboratory frame. If a single-axis detector is placed near the sample, only a cosine-modulated signal is detected with no information regarding the direction of precession. This measurement setup results in spectral crowding as all detected transitions lead to both positive and negative peaks (Fig.\,\ref{nutation_R1_R2_DD_RFF}b). To overcome these issues, we implemented a quadrature detection scheme, as described previously\cite{fabricant_proton_2024}. Thus, we are able to discriminate the absolute sign of the detected peaks as well as increase the overall SNR and simplify the spectrum. The two different configurations are exemplified in Fig.\,\ref{nutation_R1_R2_DD_RFF}b, where the simulated spectra are obtained by choosing either the `single-channel' operator $\hat{\rho}_{\text{det}}=\gamma_{I}\hat{I}_{x} + \gamma_{S}\hat{S}_{x}$ or `quadrature' operator $\hat{\rho}_{\text{det}}=\gamma_{I}\hat{I}_{+} + \gamma_{S}\hat{S}_{+}$, for detection. The maximum number of observable peaks is four, provided that the spectral separation between peaks is larger than the characteristic linewidths. We label these peaks as `near-zero-frequency' (nZF) peaks ($\nu_{13}$ and $\nu_{34}$) and $J$ peaks ($\nu_{12}$ and $\nu_{24}$). At zero field, only the zero-frequency (overlapping $\nu_{13}$ and $\nu_{34}$ peaks) and $\pm J$ peaks are observed; by increasing the main field, the splitting between energy levels increases (Fig.\,\ref{nutation_R1_R2_DD_RFF}c) such that peaks shift to higher frequencies, ultimately leading to two doublets centered around $^1$H and $^{13}$C Larmor frequencies at high field. 

In the limiting case of zero field ($\theta \rightarrow \pi/4$), we compute analytical expressions for the theoretical nutation curves of the corresponding peaks $A_{\nu_{ij}}(t_p)$ as a function of pulse duration $t_{p}$ (assuming $t_{p}\ll 1/J_{IS}$). The expressions for the nutation curves shown in Fig. \ref{nutation_R1_R2_DD_RFF}d are given by 

\begin{widetext}
\begin{align}
    A_{\nu_{13}}(t_p)  &\propto  p_{\psi_1}(t_s)(\sin{( B_{y}\gamma_{\text{C}}t_{p})} + \sin{( B_{y}\gamma_{\text{H}}t_{p}})) + p_{\psi_3}(t_s) \sin{( B_{y}(\gamma_{\text{H}} + \gamma_{\text{C}})t_{p})}\label{eq:nut13} \\
    A_{\nu_{34}}(t_p) &\propto p_{\psi_4}(t_s)(\sin{( B_{y}\gamma_{\text{C}}t_{p})} + \sin{( B_{y}\gamma_{\text{H}}t_{p}})) - p_{\psi_3}(t_s)\sin{( B_{y}(\gamma_{\text{H}} + \gamma_{\text{C}})t_{p})} \label{eq:nut34} \\
    A_{\nu_{12}}(t_p) &\propto p_{\psi_1}(t_s)(\sin{( B_{y}\gamma_{\text{C}}t_{p})} - \sin{( B_{y}\gamma_{\text{H}}t_{p}})) + p_{\psi_2}(t_s)\sin{( B_{y}(\gamma_{\text{H}} - \gamma_{\text{C}})t_{p})}\label{eq:nut12} \\
    A_{\nu_{24}}(t_p) &\propto p_{\psi_4}(t_s)(\sin{(B_{y}\gamma_{C}t_{p})} - \sin{( B_{y}\gamma_{H}t_{p}})) - p_{\psi_2}(t_s)\sin{( B_{y}(\gamma_{\text{H}} - \gamma_{\text{C}})t_{p})} \,.
    \label{eq:nut24}
\end{align}
\end{widetext}
where $p_{\psi_i}(t_s)$ represents the time-dependent population of eigenstate $\ket{\psi_i}$ after decay during a storage time $t_s$.

As the eigenstates of the Hamiltonian characteristic to both storage and measurement time are field-dependent, their initial populations subsequent to adiabatic shuttling $\{p_{\psi 1}(0),p_{\psi 2}(0),p_{\psi 3}(0),p_{\psi 4}(0) \}=\{\delta, \epsilon, -\epsilon, -\delta \}$ will decay at different rates and contribute to the observed transitions in a complex manner. The general equations describing population evolution in zero-to-ultralow fields after adiabatic shuttling for a storage time $t_{s}$, due solely to $^{13}$C-$^1$H dipole-dipole (DD) interaction, are
\begin{flalign}
&p_{\psi_{1}}(t_s)=\delta \cdot e^{-R_{1}^{\text{fast}}t_s}+\frac{\epsilon}{3}\left(e^{-R_{1}^{\text{inter}}(\theta)t_s}-e^{-R_{1}^{\text{slow}}(\theta)t_s}\right)
\label{eq:p1_decay} \\
&p_{\psi_{2}}(t_s)=\epsilon \cdot e^{-R_{1}^{\text{slow}}(\theta)t_s}
\label{eq:p2_decay} \\
&p_{\psi_{3}}(t_s)=-\frac{\epsilon}{3}\left(2e^{-R_{1}^{\text{inter}}(\theta)t_s}+e^{-R_{1}^{\text{slow}}(\theta)t_s}\right)
\label{eq:p3_decay} \\
&p_{\psi_{4}}(t_s)=-\delta \cdot e^{-R_{1}^{\text{fast}}t_s}+\frac{\epsilon}{3}\left (e^{-R_{1}^{\text{inter}}(\theta)t_s}-e^{-R_{1}^{\text{slow}}(\theta)t_s}\right )
\label{eq:p4_decay}
\end{flalign}
with $R_{1}^{\text{fast}}=\frac{15b_{IS}^{2}\tau_{c}^{\text{DD}}}{10}$, $R_{1}^{\text{inter}}(\theta)=\frac{b_{IS}^{2}\tau_{c}^{\text{DD}}}{20}(11-2\sin2\theta)(1+\sin2\theta)$ and $R_{1}^{\text{slow}}(\theta)=\frac{b_{IS}^{2}\tau_{c}^{\text{DD}}}{10}(5+2\sin2\theta)(1-\sin2\theta)$, where $b_{IS}=-\frac{\mu_{0}}{4\pi}\frac{\gamma_{I}\gamma_{S}\hbar}{r_{IS}^{3}}$ is the dipole-dipole coupling constant between the two spins at a distance $r_{IS}$, with  $\tau_{c}^{\text{DD}}$ being the rotational correlation time characterizing molecular tumbling (See Appendix~\ref{appex:relax}). These features represent a new source of relaxation dispersion compared to standard FC experiments, where changes in decay rates originate normally from the field dependence of the spectral function $J(\omega)$, given that polarization and detection are done at the same high field.

In the limiting case of zero field ($\theta \rightarrow \pi/4$), the $\psi_{2}$ state becomes the dipolar-immune singlet state $\ket{S_{0}}$. Meanwhile, the triplet populations equilibrate quickly toward thermal equilibrium (essentially equalized populations at zero field). 
Thus, assuming only dipolar interaction at zero field, the system should reach a hyperpolarized stationary state (immune to both coherent and incoherent evolution) given by the population imbalance between the singlet state and the average of the triplet manifold, known as a long-lived state\cite{pileio_long-lived_2020,emondts_long-lived_2014}. 
In practice, additional interactions such as intermolecular dipolar couplings will lead to equilibration of the singlet-triplet imbalance. These can be modeled by the Random Field Fluctuation (RFF) mechanism which also leads to multi-exponential field-dependent population dynamics during storage time similar the to dipole interaction (see Appendix \ref{appex:relax}). 

Additional information regarding relaxation mechanisms can be obtained by measuring peak linewidths, which are proportional to the auto-relaxation rates of individual coherences. At ultralow field, dipolar interactions induce differential line broadening between the inner ($\ket{\psi_{3}}\bra{\psi_{1}}$ and  $\ket{\psi_{3}}\bra{\psi_{4}}$) and outer ($\ket{\psi_{2}}\bra{\psi_{1}}$ and $\ket{\psi_{2}}\bra{\psi_{4}}$) transitions. The auto-relaxation rates of coherences connecting $\ket{\psi_{i}}\leftrightarrow\ket{\psi_j}$ eigenstates, denoted  $R_{2}^{\nu_{ij}}$, are 
\begin{flalign}
& R_{2}^{\nu_{13},\nu_{34}}(\theta)=\frac{b_{IS}^{2}\tau_{c}^{\text{DD}}}{20}(17+7\sin2\theta)\\
& R_{2}^{\nu_{12},\nu_{24}}(\theta)=\frac{b_{IS}^{2}\tau_{c}^{\text{DD}}}{20}(17-7\sin2\theta) \,.
\end{flalign}

In Fig.~\ref{nutation_R1_R2_DD_RFF}e--h, we highlight the different impact and field-scaling profile of the two relaxation mechanisms, DD and RFF, implying that no universal constant rate can be added to the rates based on the dipolar interaction in order to fit experimental data. In this work, we use a general field-dependent model to fit simultaneously the set of observable decays (see Section III in Supporting Information) with fitting variables describing molecular tumbling ($\tau_{c}^{\text{DD}},\tau_{c}^{\text{RFF}}$), internuclear distances ($r_{\text{CH}}$), local field fluctuations ($\omega_{\text{rms}}^{\text{C}},\omega_{\text{rms}}^{\text{H}}$), and their correlations ($\kappa_{\text{CH}}$).

As all detectable peaks originate from linear combinations of time-dependent populations (Eqs.\,\ref{eq:p1_decay}--\ref{eq:p4_decay}), the observed signals will display a multi-exponential decay. At the same time, the apparent peak decays can be modulated by the nutation angle induced by the detection pulse, unveiling different evolution processes during storage time. From Eqs.\,\ref{eq:nut13}--\ref{eq:nut24}, we see that a suitable choice of pulse duration $t_p$ can lead to selective observation of just one type of population, each with a different decay rate.
The model described above can be easily extended to larger spin systems like the $^{13}$CH$_3$ spin system in [$^{13}$C]-methanol (see Section V of Supporting Information).  

\begin{figure*}[htb]
\includegraphics[trim=0pt 30pt 0pt 50pt, clip, width=\textwidth]{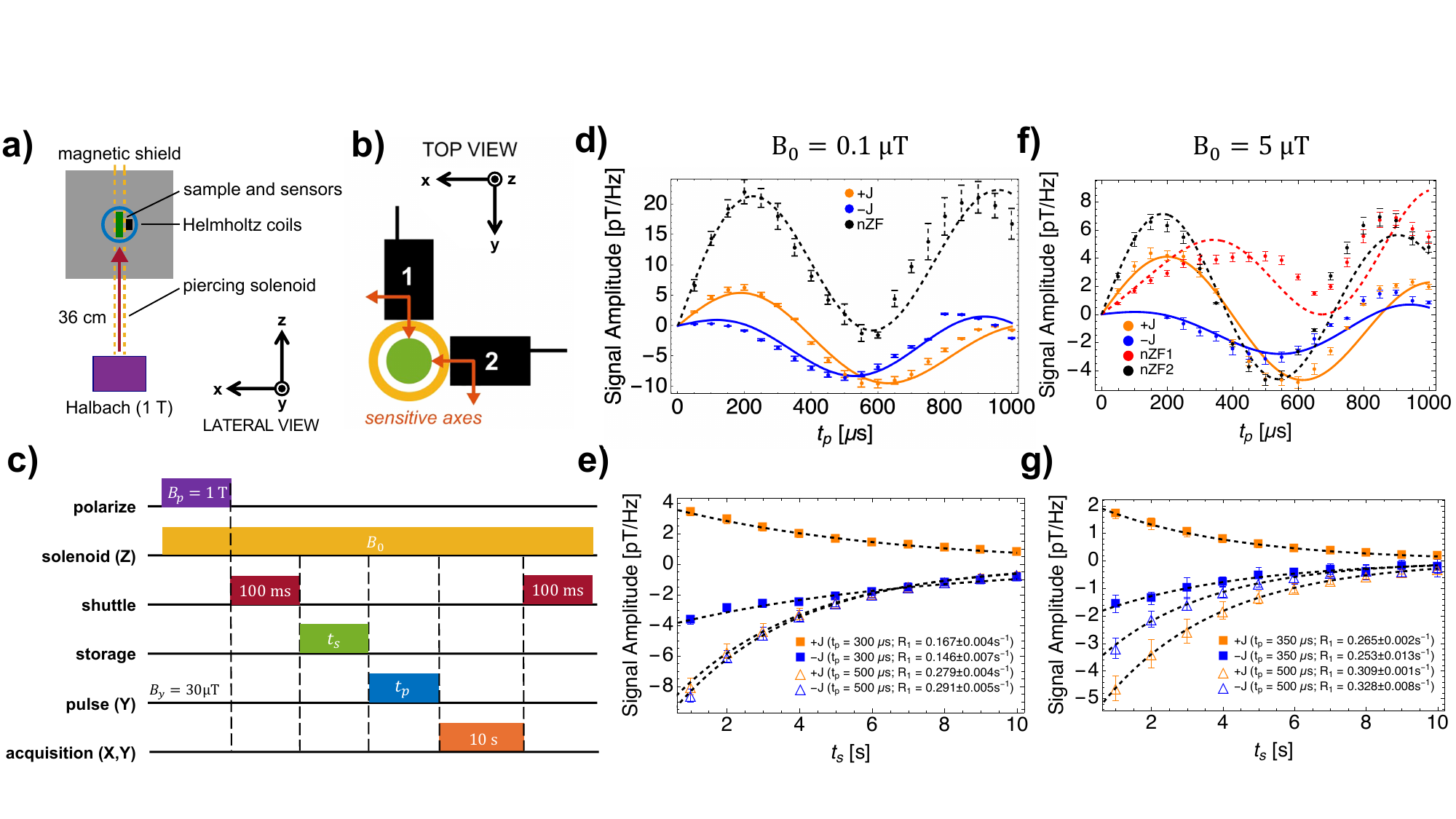}
\caption{a) Schematic representation of the experimental setup (lateral view in the laboratory frame): the sample is shuttled from a prepolarizing 1\,T Halbach magnet array through a distance of around 36\,cm in a piercing solenoid into the magnetically shielded region for storage, pulse and detection. b) Top view of gradiometric quadrature detection scheme, consisting of two optical atomic magnetometers (dual-axis vector detectors) with four acquisition channels. c) Pulse sequences used for nutation (indicated by pulse time $t_p$) and polarization-lifetime (indicated by storage time $t_s$) experiments in nT--$\upmu$T fields. Spin polarization in the Halbach array lasted 10\,s. d) Nutation curves of the three observed peaks for the [$^{13}$C] formic-acid sample, using a detection pulse with a constant background field amplitude of $B_y = -30\,\upmu$T for storage time $t_{s} = 1 s$. Equations \ref{eq:nut13}--\ref{eq:nut24} were used to fit individual nutation profiles. e) Signal decay for $+J$ and $-J$ peaks as a function of storage time $t_s$ using two different pulse lengths $t_p$ prior to detection. Decay profiles were fitted with a mono-exponential function $S(t_s)=A\exp(-R_1\,t_s)$, with characteristic rates shown as inset. All experiments were performed at a measurement field of $B_0=0.1\,\upmu$T. f) Nutation curves of the four observed peaks nZF1 ($\nu_{34}$), nZF2 ($\nu_{13}$), $+J$ ($\nu_{24}$), and $-J$ ($\nu_{12}$) and g) Signal decay for $+J$ and $-J$ peaks as a function of storage time $t_s$ for similar [$^{13}$C] formic-acid experiments performed at a measurement field of $B_0=5\,\upmu$T.
} 
\label{pulse_seq}
\end{figure*}

We now focus on the experimental results obtained for a [$^{13}$C-$^{1}$H] spin system characteristic to [$^{13}$C] formic-acid solutions. The experiments were performed with a custom setup as described in \cite{fabricant_proton_2024} and schematically represented in Fig.\,\ref{pulse_seq}a. Samples are first polarized in a 1\,T Halbach magnet and shuttled through a piercing solenoid into a shielded region for storage, pulse, and detection. We developed a gradiometric quadrature detection method which utilizes two orthogonal dual-axis magnetometers (Fig.\,\ref{pulse_seq}b), with a total of four acquisition channels from which a complex signal is constructed\cite{fabricant_proton_2024}.
Figure \ref{pulse_seq}c shows the pulse sequence used for all  experiments, where the pulse duration ($t_p$) for measuring nutation curves and the storage time ($t_s$) for extracting polarization lifetimes were varied independently. Each experiment represents the average of signals from a total of 64 automated repetitions, including shuttling from high field to low field.

As mentioned above, certain choices of nutation angles can help isolate the contributions of different populations to the observed signal. 
Figure \ref{pulse_seq}d shows the nutation curves of the three observable peaks of $^{13}$C-labeled formic acid at $B_0=0.1\,\upmu$T. Due to the small ($<1$\,mHz) frequency difference and finite linewidths, the $\nu_{13}$ and $\nu_{34}$ transitions appear as a single peak which is labeled `nZF'. The other two transitions ($\nu_{24}$ and $\nu_{12}$) are labeled `$+J$' and `$-J$', respectively. The plot shows the signal amplitude of individual peaks as a function of pulse duration, with a constant pulse field amplitude ($B_y=-30\upmu$T) applied after shuttling ($t_s=1$\,s). Equations \ref{eq:nut13}--\ref{eq:nut24} were used to fit the experimental points, showing good agreement (average $R^2>0.95$). The different nutation profiles for $+J$ and $-J$ peaks arise from the $\hat{Z}_x$ component of the density operator $\hat{\rho}_{\text{ad}}$ (Eq.\,\ref{eq:rho_adiabatic}) after shuttling (see Section II in Supporting Information). For a sudden transfer to ultralow fields, the density operator becomes $\hat{\rho}_{\text{sud}}=\frac{\delta-\epsilon}{2}\hat{I}_{z}+\frac{\delta+\epsilon}{2}\hat{S}_{z}$, and the $+J$ and $-J$ nutation profiles would therefore be different from the adiabatic case considered here.

\renewcommand{\arraystretch}{1.3} 

\begin{table}[htb]
\centering
\caption{Best-fit relaxation parameters extracted from a global fit of experimental data at $0.1\ \upmu\text{T}$ and $5\ \upmu\text{T}$ fields using the theoretical model described above (see Section III in Supporting Information for computational details).
}
\label{tab:fit}
\resizebox{\linewidth}{!}{%
\begin{tabular}{c|cc|cccc}
\multirow{2}{*}{Field ($\upmu$T)} &
\multicolumn{2}{c|}{Dipolar} &
\multicolumn{4}{c}{Random Field Fluctuations} \\
\cline{2-7}
& $d_{\text{CH}}$\,(\AA) & $\tau_{C}^{\text{DD}}$\,(ps) & $\omega_{\text{rms}}^{\text{H}}$\,(rad) & $\omega_{\text{rms}}^{\text{C}}$\,(rad) & $\kappa_{\text{CH}}$ & $\tau_{c}^{\text{RFF}}$\,(ps) \\
\hline
0.1\,$\upmu$T \& 5\,$\upmu$T  & 1.1\footnotemark[1] & 3.24 & $49.63\cdot10^{3}$  & $85.32\cdot10^{3}$ & 0.49 & 19.34 \\
\hline
\end{tabular}%
}
\footnotetext[1]{Fixed value taken from literature.}
\end{table}

Thus, we have further evidence (in addition to the simulated spin dynamics during field shuttling from Fig. S1) based on the nutation data that our shuttling was performed adiabatically, which is an essential aspect for properly analyzing polarization lifetimes in ultralow fields (See Sections II and VI in Supporting Information where we consider the case of sudden transfer in more detail).

We next performed polarization-lifetime measurements by increasing the storage time before the pulse ($t_s \in [0;10\,\text{s}]$) and tracked individual peak decays using different pulse durations ($t_p \in [0; 1 \,\text{ms}]$). Firstly, we observed that the measured decay rates depended strongly on the nutation angle used for observation (Fig.\,\ref{pulse_seq}e). This phenomenon stems from the modulation of each individual population's contribution to the observed signal: for $t_p=300\,\upmu$s both $+J$ and $-J$ peaks originate from the $\psi_{2}$ population (Eqs.\,\ref{eq:nut12}--\ref{eq:nut24}) which has a slow decay (Eq.\,\ref{eq:p2_decay}), while at $t_p=500\,\upmu$s the two peaks come from the populations of $\psi_4$ and $\psi_1$, respectively, both having enhanced relaxation (Eqs.\,\ref{eq:p1_decay} and \ref{eq:p4_decay}). Thus, the apparent decay rate can be experimentally tuned by adjusting the pulse length (or amplitude), targeting desired features such as extending storage time before detection. This approach may be relevant for extended pulse sequences as well as experiments featuring hyperpolarized samples, which require chemical or physical manipulation prior to signal acquisition\cite{van_dyke_relayed_2022}.

Furthermore, we conducted both nutation and polarization lifetime experiments at $B_0 = 5\,\upmu$T (Fig.\,\ref{nutation_R1_R2_DD_RFF}f--g) and observed a similar trend in decay rates at two different pulse lengths. The extracted rates characteristic to the $\psi_2$ population decay (for $t_p = 350\,\upmu$s) are significantly higher than the corresponding ones at $B_0=0.1\,\upmu$T, as predicted by theory (Fig.\,\ref{nutation_R1_R2_DD_RFF}e). 

In order to derive relevant molecular parameters such as dipolar couplings, local field fluctuations, or tumbling rates, we performed a global fit on all datasets combining nutation and decay profiles at both $0.1\,\upmu$T and $5\,\upmu$T fields (see Section III of Supporting Information). The extracted parameters are listed in Table\,\ref{tab:fit}. Using this set of parameters, we predict, based on our model, the relaxation rates of the coherences generating $+J$ and $-J$ peaks and compared them with experimental values derived from linewidth fitting at $B_0=0.1\,\upmu$T. The expected values of the relaxation rate for the two transitions (due to the additive contributions of DD and RFF mechanisms) are $R_{2,\pm J}^{\text{model}}\approx0.27 $\,s$^{-1}$, smaller than the experimentally measured ones of $R_{2,+J}^{*,\text{exp}}=0.405 \pm0.003 $\,s$^{-1}$ and $R_{2,-J}^{*,\text{exp}}=0.392 \pm 0.005 $\,s$^{-1}$. This mismatch can be ascribed to  field inhomogeneities detected in our experimental setup, in particular, gradients produced by the solenoid as well as by the internal compensation coils of the zero-field magnetometers used for detection (see Section IV of Supporting Information).

In summary, we highlight in this work the counterintuitive relaxation effects that occur at ultralow field, where measurements can only be performed with quantum magnetometers, not with classical induction coils. The presented theoretical model provides a framework by which the observed effects can be described in terms of  molecular parameters. Importantly, we have highlighted a rather unexpected feature of relaxation rate measurements at ultralow field: the nutation angle induced by the detection pulse modulates the apparent relaxation rates measured through the individual peaks, thereby providing additional contrast, spectral editing, and spectral assignment mechanisms.

In the ULF regime, spectra reach their maximum complexity in terms of the total number of individual peaks\cite{appelt_paths_2010}, each decaying differently. This effect makes the ULF regime richer in information than the high-field regime, as the same relaxation mechanism can be tracked with multiple different observables. Potentially, this process could allow for a better quantification  of each relaxation mechanism's contribution. Close to zero field, differences in the precession frequencies in the laboratory frame become of the same order as the relaxation rates, such that the `secular approximation' is not applicable anymore. Furthermore, relaxation processes can also be incorporated into magnetic resonance imaging experiments, thereby producing innovating image contrast.
Finally, experiments performed under ZULF conditions are compatible with examining samples enclosed in (thick) conductive casings, such as, for example, batteries\cite{fabricant_enabling_2025}. Monitoring relaxation rates in batteries could be used to report on the dynamic processes in functioning devices.

\textit{Acknowledgment}--A.J. acknowledges funding from the US National Science Foundation, award no. CHE 2505792, and an award from the ACS PRF 68117-ND6. F.T. acknowledges funding from National Medical Project MySMIS 326475 and Project ELI-RO/RDI/2024/14 SPARC funded by the Institute of Atomic Physics (Romania). This work was supported in part by the Deutsche Forschungsgemeinschaft (DFG,German Research Foundation) in the framework to the collaborative research center ``Defects and Defect Engineering in Soft Matter''
(SFB1552) under Project No. 465145163 and under the DFG/ANR grant BU 3035/24-1.  

\appendix 

\section{Eigenstates at ultralow fields}
\label{appenx:eigenstates}
The master equation describing both coherent and incoherent evolution of the density operator $\hat{\rho}(t)$ in Bloch-Redfield-Wangsness  theory has the following form:

\begin{equation}
    \frac{\partial\hat{\rho}(t)}{\partial t} = -i\left[\hat{H}_{0},\hat{\rho}(t)-\hat{\rho}_{\text{eq}}\right] + \hat{\hat{\Gamma}}\left(\hat{\rho}(t)-\hat{\rho}_{\text{eq}}\right) \,.
    \label{eq:Lioville_Equation}
\end{equation}
Here, the static Hamiltonian $\hat{H}_{0}$ determines the coherent evolution experienced by all spins in the sample, which for molecules in isotropic liquids amounts to the Zeeman and scalar-coupling interactions. The second term represents the relaxation superoperator $\hat{\hat{\Gamma}}$ driving the system towards the thermal equilibrium density operator $\hat{\rho}_{\text{eq}}$.

For a heteronuclear two-spin-1/2 \textit{\{I, S\}} system characterized by $\gamma_{I}$ and $\gamma_{S}$ gyromagnetic ratios (in $\text{rad} \cdot \text{s}^{-1} \cdot $T$^{-1}$) and a scalar coupling constant $J_{IS}$ (in Hz), the Hamiltonian describing the interaction of this system with a magnetic field $B_0$ takes the form
\begin{equation}
    \hat{H}_{0} = -\left(\gamma_{I}\hat{I}_{z}+\gamma_{S}\hat{S}_{z}\right)B_{0} + 2\pi J_{IS} \boldsymbol{\hat{I}}\cdot\boldsymbol{\hat{S}} \,,
    \label{eq:Total_Hamiltonian}
\end{equation}
with the eigenstates 
\begin{flalign}
& \psi_{1}=\ket{\alpha\alpha}\\
& \psi_{2} = -\sin{\theta}\ket{\alpha\beta} + \cos{\theta}\ket{\beta\alpha}\\
& \psi_{3} = \cos{\theta}\ket{\alpha\beta} + \sin{\theta}\ket{\beta\alpha}\\
& \psi_{4}=\ket{\beta\beta}
\label{eq:eigenstates}
\end{flalign}
and their eigenvalues
\begin{flalign}
&E_{1} = \frac{1}{2}\left((\gamma_{I}+\gamma_{S})B_{0}+ \pi J_{IS}\right)\\
&E_{2}  = -\frac{1}{2}\left(\sqrt{((\gamma_{I}-\gamma_{S})B_{0})^{2}+(2\pi J_{IS})^{2}} +\pi J_{IS}\right)\\
& E_{3}  = \frac{1}{2}\left(\sqrt{((\gamma_{I}-\gamma_{S})B_{0})^{2}+(2\pi J_{IS})^{2}} -\pi J_{IS}\right)\\
& E_{4} = \frac{1}{2}\left (-(\gamma_{I}+\gamma_{S})B_{0} + \pi J_{IS}\right ) \,.
\label{energy_levels}
\end{flalign}

Two limiting cases are relevant for our discussion: in the weakly coupled regime ($\theta \rightarrow 0$), the system is described by the Zeeman basis $\boldsymbol{Q_{Z}}=\{\ket{\alpha\alpha},\ket{\alpha\beta},\ket{\beta\alpha},\ket{\beta\beta}\}$, while in the strongly coupled regime ($\theta \rightarrow \frac{\pi}{4}$), the eigenbasis is the singlet-triplet basis $\boldsymbol{Q_{ST}} = \{\ket{\alpha\alpha},\frac{1}{\sqrt{2}}(\ket{\alpha\beta}-\ket{\beta\alpha}),\frac{1}{\sqrt{2}}(\ket{\alpha\beta}+\ket{\beta\alpha}),\ket{\beta\beta}\}$. 

Because the singlet state $\ket{S_{0}}=\frac{1}{\sqrt{2}}\left(\ket{\alpha\beta}-\ket{\beta\alpha}\right)$ is antisymmetric with respect to the permutation operator ($\hat{P}_{\alpha\leftrightarrow\beta}\ket{S_{0}}=-\ket{S_{0}}$), while the other triplet states are symmetric ($\hat{P}_{\alpha\leftrightarrow\beta}\ket{T_{\mu}}=\ket{T_{\mu}}$ with $\mu \in \{-1,0,1\}$), no symmetric interaction like the dipolar coupling between \textit{I} and \textit{S} can mix these states\cite{pileio_long-lived_2020}. Moreover, the singlet-state population is known to be immune to the intra-pair dipolar interaction and therefore has an extended lifetime\cite{carravetta_beyond_2004}. However, the Zeeman states follow different transformations under permutation, and no population is immune to the dipolar coupling. These previously known results suggest that the relaxation rates of spin-state populations and coherences are directly impacted by the structure of the eigenstates and, thus, by the field at which storage and detection are performed. 

\section{Relaxation Dynamics during Storage Time}
\label{appex:relax}
The non-equilibrium polarization following adiabatic shuttling commutes with the main Hamiltonian $\hat{H}_{0}$, meaning that detection pulses are needed to detect spin-state transitions. At ZULF, where measurements are performed, the equilibrium polarization is negligible. Thus, the (hyper)polarized nuclear spin order will gradually decay until a new equilibrium (near-zero-polarization) state is achieved. The irreversible polarization loss is caused by various relaxation mechanisms via random molecular tumbling that induces spin transitions.

In liquid-state NMR, the dominant interactions that lead to loss of spin order are intramolecular dipole-dipole (DD) interactions, chemical shift anisotropy (CSA), and electric-field-gradient (EFG) interactions with quadrupolar nuclei\cite{kowalewski_nuclear_2019}.
For our heteronuclear two-spin-1/2 system at ultralow field, the CSA relaxation contribution is negligible as it scales as $B_{0}^{2}$\cite{kowalewski_nuclear_2019}. At the same time, EFGs only impact nuclei with spin greater or equal to 1. Some other mechanisms (e.g. intermolecular dipolar couplings\cite{hwang_dynamic_1975} or spin-rotation interaction\cite{singer_nmr_2018}) or unknown relaxation sources can be further modeled by random-field fluctuations (RFF). In the fast-tumbling approximation, the relaxation superoperator describing the intramolecular dipolar interaction is\cite{levitt_spin_2013}

\begin{equation}
    \hat{\hat{\Gamma}}_{\text{DD}}=-\frac{6}{5}b_{IS}^{2}\tau_{c}^{\text{DD}}\sum_{m=-2}^{2}(-1)^{m}\left [\hat{T}_{2m},[\hat{T}_{2-m},\cdot\,]\right ]\,,
    \label{eq:RDD_relax}
\end{equation}
where $\hat{T}_{rm}$ is the spherical tensor operator of rank \textit{r} and order \textit{m}. The fast-tumbling approximation is adequate in ultralow fields, as the spectral density $J(\omega)\propto \frac{\tau_{c}^{\text{DD}}}{1+(\omega\tau_{c}^{\text{DD}})^{2}}$ will always be field-independent ($\omega\tau_{c}^{\text{DD}}\ll1$). Field-dependent rates describing populations' auto- and cross-relaxation are computed from Liouville products\cite{cavanagh_protein_1995} of the form 

\begin{equation}
    \Gamma_{ij}=\frac{\braket{\ket{\psi_i}\bra{\psi_i}|\hat{\hat{\Gamma}} |\ket{\psi_j}\bra{\psi_j}}}{\braket{\ket{\psi_i}\bra{\psi_i}|\ket{\psi_i}\bra{\psi_i}}}\,, i,j \in \{1, 2, 3, 4\} \,.
    \label{eq:Liouville_product_pop}
\end{equation}

Using these $\Gamma_{ij}$ elements, we build the matrix representation of the system of coupled differential equations describing both population decay and mixing in our system during the storage time:

\begin{widetext}
\begin{align}
 \frac{d}{dt}\begin{pmatrix}
p_{\psi1}(t)  \\ p_{\psi2}(t)  \\p_{\psi3}(t)  \\p_{\psi4}(t)  
\end{pmatrix}   =
\frac{b_{IS}^{2}\tau_{c}^{\text{DD}}}{20}
\begin{pmatrix}
-18 & 3(1-\sin2\theta) &3(1+\sin2\theta) &12\\
3(1-\sin2\theta) &2(4+\sin2\theta)(1-\sin2\theta) & 2(1-\sin2\theta)(1+\sin2\theta)& 3(1-\sin2\theta)\\
3(1+\sin2\theta) & 2(1+\sin2\theta)(1-\sin2\theta) &2(-4+\sin2\theta)(1+\sin2\theta)  &3(1+\sin2\theta)\\
12 & 3(1-\sin2\theta) &3(1+\sin2\theta) &-18
\end{pmatrix}
\begin{pmatrix}
p_{\psi1}(t) \\ p_{\psi2}(t)  \\p_{\psi3}(t) \\p_{\psi4}(t)
\label{eq:matrix_eq}
\end{pmatrix}\,,
\end{align}
\end{widetext}
whose solutions are given by Eqs.~\ref{eq:p1_decay}--\ref{eq:p4_decay}.

The RFF mechanism can be invoked to describe the effects of other mechanisms that are often difficult to quantify, such as the intermolecular dipolar coupling or the spin-rotation interaction\cite{whipham_cross-correlated_2024}. Similarly to Eq.\,\ref{eq:RDD_relax}, we can build the corresponding relaxation superoperator for RFF as 
\begin{equation}
\hat{\hat{\Gamma}}_{\text{RFF}}=\sum_{i,j=1}^{2}\kappa_{ij}\omega_{\text{rms}}^{i}\omega_{\text{rms}}^{j}\tau_{c}^{\text{RFF}}\sum_{m=-1}^{1}(-1)^{m}[\hat{T}_{1m},[\hat{T}_{1-m},\cdot\,]]
\end{equation}
with $\tau_{c}^{\text{RFF}}$ being the correlation time of the RFF and $\omega_{\text{rms}}^{i,j}$ being the root-mean-square amplitudes of the local field fluctuations associated with spins \textit{i} and \textit{j}; $-1\leq\kappa_{ij}\leq1$ is the coefficient describing the degree of correlation of fluctuations experienced by the two spins. By definition, $\kappa_{ii}=\kappa_{jj}=1$, while $\kappa_{ij}=\pm1$ represents a perfectly correlated ($+1$) or anti-correlated ($-1$) interaction between the two spins. The system of coupled differential equations describing population dynamics during storage time due to RFF is similar to the one in Eq. \ref{eq:matrix_eq} and has an identical solution. A Mathematica notebook summarizing step-by-step derivations of spin dynamics and analytical expressions of relaxation rates is attached as Supporting Information. 

\bibliography{references}

@misc{fabricant_enabling_2025,
	title = {Enabling nondestructive observation of electrolyte composition in batteries with ultralow-field nuclear magnetic resonance},
	url = {https://chemrxiv.org/engage/chemrxiv/article-details/6773ed7d6dde43c908e492eb},
	doi = {10.26434/chemrxiv-2024-32xj9-v3},
	urldate = {2025-01-07},
	publisher = {ChemRxiv},
	author = {Fabricant, Anne M. and Picazo-Frutos, Román and Teleanu, Florin and Rees, Gregory J. and Lin, Mengjiang and House, Robert A. and Bruce, Peter G. and Blanchard, John and Eills, James and Sheberstov, Kirill and Budker, Dmitry and Barskiy, Danila A. and Jerschow, Alexej},
	month = jan,
	year = {2025},
}

@book{cavanagh_protein_1995,
	title = {Protein {NMR} {Spectroscopy}: {Principles} and {Practice}},
	isbn = {978-0-08-051529-8},
	shorttitle = {Protein {NMR} {Spectroscopy}},
	publisher = {Elsevier},
	author = {Cavanagh, John and Fairbrother, Wayne J. and III, Arthur G. Palmer and Skelton, Nicholas J.},
	month = nov,
	year = {1995},
	note = {Google-Books-ID: 85rYGWiBJ1kC},
}

@article{appelt_paths_2010,
	title = {Paths from weak to strong coupling in {NMR}},
	volume = {81},
	url = {https://link.aps.org/doi/10.1103/PhysRevA.81.023420},
	doi = {10.1103/PhysRevA.81.023420},
	number = {2},
	urldate = {2024-01-22},
	journal = {Physical Review A},
	author = {Appelt, S. and Häsing, F. W. and Sieling, U. and Gordji-Nejad, A. and Glöggler, S. and Blümich, B.},
	month = feb,
	year = {2010},
	pages = {023420},
}

@article{stern_simulation_2023,
	title = {Simulation of {NMR} spectra at zero and ultralow fields from {A} to {Z} – a tribute to {Prof}. {Konstantin} {L}'vovich {Ivanov}},
	volume = {4},
	url = {https://mr.copernicus.org/articles/4/87/2023/},
	doi = {10.5194/mr-4-87-2023},
	number = {1},
	urldate = {2023-04-17},
	journal = {Magnetic Resonance},
	author = {Stern, Quentin and Sheberstov, Kirill},
	month = apr,
	year = {2023},
	pages = {87--109},
}

@book{kuprov_spin_2023,
	title = {Spin: {From} {Basic} {Symmetries} to {Quantum} {Optimal} {Control}},
	isbn = {978-3-031-05607-9},
	shorttitle = {Spin},
	publisher = {Springer Nature},
	author = {Kuprov, Ilya},
	month = mar,
	year = {2023},
	note = {Google-Books-ID: edmzEAAAQBAJ},
}

@article{stenstrom_how_2022,
	title = {How does it really move? {Recent} progress in the investigation of protein nanosecond dynamics by {NMR} and simulation},
	volume = {77},
	issn = {0959-440X},
	shorttitle = {How does it really move?},
	url = {https://www.sciencedirect.com/science/article/pii/S0959440X22001385},
	doi = {10.1016/j.sbi.2022.102459},
	urldate = {2022-10-09},
	journal = {Current Opinion in Structural Biology},
	author = {Stenström, Olof and Champion, Candide and Lehner, Marc and Bouvignies, Guillaume and Riniker, Sereina and Ferrage, Fabien},
	month = dec,
	year = {2022},
	pages = {102459},
}

@article{charlier_protein_2016,
	title = {Protein dynamics from nuclear magnetic relaxation},
	volume = {45},
	issn = {1460-4744},
	url = {https://pubs.rsc.org/en/content/articlelanding/2016/cs/c5cs00832h},
	doi = {10.1039/C5CS00832H},
	number = {9},
	urldate = {2020-11-25},
	journal = {Chemical Society Reviews},
	author = {Charlier, Cyril and Cousin, Samuel F. and Ferrage, Fabien},
	month = may,
	year = {2016},
	pages = {2410--2422},
}

@article{van_dyke_relayed_2022,
	title = {Relayed hyperpolarization for zero-field nuclear magnetic resonance},
	volume = {8},
	url = {https://www.science.org/doi/10.1126/sciadv.abp9242},
	doi = {10.1126/sciadv.abp9242},
	number = {29},
	urldate = {2023-04-17},
	journal = {Science Advances},
	author = {Van Dyke, Erik T. and Eills, James and Picazo-Frutos, Román and Sheberstov, Kirill F. and Hu, Yinan and Budker, Dmitry and Barskiy, Danila A.},
	month = jul,
	year = {2022},
	pages = {eabp9242},
}

@incollection{harris_zero-_2016,
	address = {Chichester, UK},
	title = {Zero- to {Ultralow}-{Field} {NMR}},
	isbn = {978-0-470-03459-0 978-0-470-05821-3},
	url = {http://doi.wiley.com/10.1002/9780470034590.emrstm1369},
	urldate = {2023-07-21},
	booktitle = {{eMagRes}},
	publisher = {John Wiley \& Sons, Ltd},
	author = {Blanchard, John W. and Budker, Dmitry},
	editor = {Harris, Robin K. and Wasylishen, Roderick L.},
	month = sep,
	year = {2016},
	doi = {10.1002/9780470034590.emrstm1369},
	pages = {1395--1410},
}

@article{carravetta_beyond_2004,
	title = {Beyond the {T1} {Limit}: {Singlet} {Nuclear} {Spin} {States} in {Low} {Magnetic} {Fields}},
	volume = {92},
	shorttitle = {Beyond the \$\{{T}\}\_\{1\}\$ {Limit}},
	url = {https://link.aps.org/doi/10.1103/PhysRevLett.92.153003},
	doi = {10.1103/PhysRevLett.92.153003},
	number = {15},
	urldate = {2020-03-29},
	journal = {Physical Review Letters},
	author = {Carravetta, Marina and Johannessen, Ole G. and Levitt, Malcolm H.},
	month = apr,
	year = {2004},
	pages = {153003},
}

@book{pileio_long-lived_2020,
	title = {Long-lived {Nuclear} {Spin} {Order}: {Theory} and {Applications}},
	isbn = {978-1-78801-568-4},
	shorttitle = {Long-lived {Nuclear} {Spin} {Order}},
	publisher = {Royal Society of Chemistry},
	author = {Pileio, Giuseppe},
	month = mar,
	year = {2020},
	note = {Google-Books-ID: 2xvdDwAAQBAJ},
}

@article{pileio_singlet_2017,
	title = {Singlet {NMR} methodology in two-spin-1/2 systems},
	volume = {98-99},
	issn = {0079-6565},
	url = {http://www.sciencedirect.com/science/article/pii/S0079656516300279},
	doi = {10.1016/j.pnmrs.2016.11.002},
	urldate = {2020-03-03},
	journal = {Progress in Nuclear Magnetic Resonance Spectroscopy},
	author = {Pileio, Giuseppe},
	month = feb,
	year = {2017},
	pages = {1--19},
}

@article{zhukov_assessment_2019,
	title = {Assessment of heteronuclear long-lived states at ultralow magnetic fields},
	volume = {21},
	issn = {1463-9084},
	url = {https://pubs.rsc.org/en/content/articlelanding/2019/cp/c9cp03719e},
	doi = {10.1039/C9CP03719E},
	number = {33},
	urldate = {2020-07-16},
	journal = {Physical Chemistry Chemical Physics},
	author = {Zhukov, Ivan V. and Kiryutin, Alexey S. and Yurkovskaya, Alexandra V. and Ivanov, Konstantin L.},
	month = aug,
	year = {2019},
	pages = {18188--18194},
}

@article{zhukov_field-cycling_2018,
	title = {Field-cycling {NMR} experiments in an ultra-wide magnetic field range: relaxation and coherent polarization transfer},
	volume = {20},
	issn = {1463-9084},
	shorttitle = {Field-cycling {NMR} experiments in an ultra-wide magnetic field range},
	url = {https://pubs.rsc.org/en/content/articlelanding/2018/cp/c7cp08529j},
	doi = {10.1039/C7CP08529J},
	number = {18},
	urldate = {2024-04-29},
	journal = {Physical Chemistry Chemical Physics},
	author = {Zhukov, Ivan V. and Kiryutin, Alexey S. and Yurkovskaya, Alexandra V. and Grishin, Yuri A. and Vieth, Hans-Martin and Ivanov, Konstantin L.},
	month = may,
	year = {2018},
	pages = {12396--12405},
}

@article{alcicek_zero-_2023,
	title = {Zero- to low-field relaxometry of chemical and biological fluids},
	volume = {6},
	copyright = {2023 Springer Nature Limited},
	issn = {2399-3669},
	url = {https://www.nature.com/articles/s42004-023-00965-8},
	doi = {10.1038/s42004-023-00965-8},
	number = {1},
	urldate = {2023-09-22},
	journal = {Communications Chemistry},
	author = {Alcicek, Seyma and Put, Piotr and Kubrak, Adam and Alcicek, Fatih Celal and Barskiy, Danila and Gloeggler, Stefan and Dybas, Jakub and Pustelny, Szymon},
	month = aug,
	year = {2023},
	pages = {1--9},
}

@article{bodenstedt_fast-field-cycling_2021,
	title = {Fast-field-cycling ultralow-field nuclear magnetic relaxation dispersion},
	volume = {12},
	copyright = {2021 The Author(s)},
	issn = {2041-1723},
	url = {https://www.nature.com/articles/s41467-021-24248-9},
	doi = {10.1038/s41467-021-24248-9},
	number = {1},
	urldate = {2025-01-07},
	journal = {Nature Communications},
	author = {Bodenstedt, Sven and Mitchell, Morgan W. and Tayler, Michael C. D.},
	month = jun,
	year = {2021},
	pages = {4041},
}

@article{budker_optical_2007,
	title = {Optical magnetometry},
	volume = {3},
	copyright = {2007 Springer Nature Limited},
	issn = {1745-2481},
	url = {https://www.nature.com/articles/nphys566},
	doi = {10.1038/nphys566},
	number = {4},
	urldate = {2025-02-10},
	journal = {Nature Physics},
	author = {Budker, Dmitry and Romalis, Michael},
	month = apr,
	year = {2007},
	pages = {227--234},
}

@article{fabricant_proton_2024,
	title = {Proton relaxometry of tree leaves at hypogeomagnetic fields},
	volume = {15},
	issn = {1664-462X},
	url = {https://www.frontiersin.org/journals/plant-science/articles/10.3389/fpls.2024.1352282/full},
	doi = {10.3389/fpls.2024.1352282},
	urldate = {2025-01-07},
	journal = {Frontiers in Plant Science},
	author = {Fabricant, Anne M. and Put, Piotr and Barskiy, Danila A.},
	month = mar,
	year = {2024},
}

@article{bolik-coulon_comprehensive_2023,
	title = {Comprehensive analysis of relaxation decays from high-resolution relaxometry},
	volume = {355},
	issn = {1090-7807},
	url = {https://www.sciencedirect.com/science/article/pii/S1090780723001908},
	doi = {10.1016/j.jmr.2023.107555},
	urldate = {2023-09-18},
	journal = {Journal of Magnetic Resonance},
	author = {Bolik-Coulon, Nicolas and Zachrdla, Milan and Bouvignies, Guillaume and Pelupessy, Philippe and Ferrage, Fabien},
	month = sep,
	year = {2023},
	pages = {107555},
}

@book{kimmich_field-cycling_2018,
	title = {Field-cycling {NMR} {Relaxometry}: {Instrumentation}, {Model} {Theories} and {Applications}},
	isbn = {978-1-78801-550-9},
	shorttitle = {Field-cycling {NMR} {Relaxometry}},
	publisher = {Royal Society of Chemistry},
	author = {Kimmich, Rainer},
	month = oct,
	year = {2018},
	note = {Google-Books-ID: MbarDwAAQBAJ},
}

@article{kumar_cross-correlations_2000,
	title = {Cross-correlations in {NMR}},
	volume = {37},
	issn = {0079-6565},
	url = {https://www.sciencedirect.com/science/article/pii/S0079656500000236},
	doi = {10.1016/S0079-6565(00)00023-6},
	number = {3},
	urldate = {2023-04-17},
	journal = {Progress in Nuclear Magnetic Resonance Spectroscopy},
	author = {Kumar, Anil and Christy Rani Grace, R. and Madhu, P. K.},
	month = sep,
	year = {2000},
	pages = {191--319},
}

@article{kumar_two-dimensional_1980,
	title = {A two-dimensional nuclear {Overhauser} enhancement ({2D} {NOE}) experiment for the elucidation of complete proton-proton cross-relaxation networks in biological macromolecules},
	volume = {95},
	issn = {0006-291X},
	url = {https://www.sciencedirect.com/science/article/pii/0006291X80906956},
	doi = {10.1016/0006-291X(80)90695-6},
	number = {1},
	urldate = {2023-04-11},
	journal = {Biochemical and Biophysical Research Communications},
	author = {Kumar, Anil and Ernst, R. R. and Wüthrich, K.},
	month = jul,
	year = {1980},
	pages = {1--6},
}

@book{noggle_nuclear_1971,
	title = {The {Nuclear} {Overhauser} {Effect}: {Chemical} {Applications}},
	isbn = {978-0-12-520650-1},
	shorttitle = {The {Nuclear} {Overhauser} {Effect}},
	publisher = {Academic Press},
	author = {Noggle, Joseph H. and Schirmer, Roger E.},
	year = {1971},
}

@article{overhauser_polarization_1953,
	title = {Polarization of {Nuclei} in {Metals}},
	volume = {92},
	url = {https://link.aps.org/doi/10.1103/PhysRev.92.411},
	doi = {10.1103/PhysRev.92.411},
	number = {2},
	urldate = {2023-04-11},
	journal = {Physical Review},
	author = {Overhauser, Albert W.},
	month = oct,
	year = {1953},
	pages = {411--415},
}

@book{kowalewski_nuclear_2019,
	title = {Nuclear {Spin} {Relaxation} in {Liquids}: {Theory}, {Experiments}, and {Applications}, {Second} {Edition}},
	isbn = {978-0-367-89006-3},
	shorttitle = {Nuclear {Spin} {Relaxation} in {Liquids}},
	publisher = {CRC Press LLC},
	author = {Kowalewski, Jozef and Maler, Lena},
	month = dec,
	year = {2019},
	note = {Google-Books-ID: BU0dzAEACAAJ},
}

@book{levitt_spin_2013,
	title = {Spin {Dynamics}: {Basics} of {Nuclear} {Magnetic} {Resonance}},
	isbn = {978-1-118-68184-8},
	shorttitle = {Spin {Dynamics}},
	publisher = {John Wiley \& Sons},
	author = {Levitt, Malcolm H.},
	month = may,
	year = {2013},
	note = {Google-Books-ID: bysFAa4MPQcC},
}

@book{abragam_principles_1961,
	title = {The {Principles} of {Nuclear} {Magnetism}},
	isbn = {978-0-19-852014-6},
	publisher = {Clarendon Press},
	author = {Abragam, Anatole},
	year = {1961},
	note = {Google-Books-ID: 9M8U\_JK7K54C},
}

@article{barskiy_zero-_2025,
	title = {Zero- to ultralow-field nuclear magnetic resonance},
	issn = {0079-6565},
	url = {https://www.sciencedirect.com/science/article/pii/S0079656525000020},
	doi = {10.1016/j.pnmrs.2025.101558},
	urldate = {2025-02-18},
	journal = {Progress in Nuclear Magnetic Resonance Spectroscopy},
	author = {Barskiy, Danila A. and Blanchard, John W. and Budker, Dmitry and Eills, James and Pustelny, Szymon and Sheberstov, Kirill F. and Tayler, Michael C. D. and Trabesinger, Andreas H.},
	month = feb,
	year = {2025},
	pages = {101558},
}

@article{singer_nmr_2018,
	title = {{NMR} spin-rotation relaxation and diffusion of methane},
	volume = {148},
	issn = {0021-9606},
	url = {https://doi.org/10.1063/1.5027097},
	doi = {10.1063/1.5027097},
	number = {20},
	urldate = {2024-11-21},
	journal = {The Journal of Chemical Physics},
	author = {Singer, P. M. and Asthagiri, D. and Chapman, W. G. and Hirasaki, G. J.},
	month = may,
	year = {2018},
	pages = {204504},
}

@article{hwang_dynamic_1975,
	title = {Dynamic effects of pair correlation functions on spin relaxation by translational diffusion in liquids},
	volume = {63},
	issn = {0021-9606},
	url = {https://doi.org/10.1063/1.431841},
	doi = {10.1063/1.431841},
	number = {9},
	urldate = {2024-10-10},
	journal = {The Journal of Chemical Physics},
	author = {Hwang, Lian‐Pin and Freed, Jack H.},
	month = nov,
	year = {1975},
	pages = {4017--4025},
}

@article{emondts_long-lived_2014,
	title = {Long-{Lived} {Heteronuclear} {Spin}-{Singlet} {States} in {Liquids} at a {Zero} {Magnetic} field},
	volume = {112},
	url = {https://link.aps.org/doi/10.1103/PhysRevLett.112.077601},
	doi = {10.1103/PhysRevLett.112.077601},
	number = {7},
	urldate = {2020-03-29},
	journal = {Physical Review Letters},
	author = {Emondts, M. and Ledbetter, M. P. and Pustelny, S. and Theis, T. and Patton, B. and Blanchard, J. W. and Butler, M. C. and Budker, D. and Pines, A.},
	month = feb,
	year = {2014},
	note = {Number: 7
Publisher: American Physical Society},
	pages = {077601},
}

@article{whipham_cross-correlated_2024,
	title = {Cross-correlated relaxation in the {NMR} of near-equivalent spin pairs: {Longitudinal} relaxation and long-lived singlet order},
	volume = {161},
	issn = {0021-9606},
	shorttitle = {Cross-correlated relaxation in the {NMR} of near-equivalent spin pairs},
	url = {https://doi.org/10.1063/5.0213997},
	doi = {10.1063/5.0213997},
	number = {1},
	urldate = {2024-07-05},
	journal = {The Journal of Chemical Physics},
	author = {Whipham, James W. and Sabba, Mohamed and Dagys, Laurynas and Moustafa, Gamal and Bengs, Christian and Levitt, Malcolm H.},
	month = jul,
	year = {2024},
	pages = {014112},
}

@article{jerschow_nuclear_2005,
	title = {From nuclear structure to the quadrupolar {NMR} interaction and high-resolution spectroscopy},
	volume = {46},
	issn = {0079-6565},
	url = {https://www.sciencedirect.com/science/article/pii/S0079656504000615},
	doi = {10.1016/j.pnmrs.2004.12.001},
	number = {1},
	urldate = {2024-03-23},
	journal = {Progress in Nuclear Magnetic Resonance Spectroscopy},
	author = {Jerschow, Alexej},
	month = mar,
	year = {2005},
	pages = {63--78},
}

\end{document}


\setstretch{1.0}
\title{Supporting Information for:\\
\textit{Nuclear spin relaxation in zero- to ultralow-field magnetic resonance spectroscopy}}

\author{Florin Teleanu}
\affiliation{Department of Chemistry, New York University, New York, NY 10003, United States}
\affiliation{ELI-NP, “Horia Hulubei” National Institute for Physics and Nuclear Engineering, 30 Reactorului Street, Bucharest-Magurele, 077125, Ilfov, Romania}
\author{Anne M. Fabricant}
\altaffiliation{Current address: Department of Biosignals, Physikalisch-Technische Bundesanstalt (PTB), 10587 Berlin, Germany}
\affiliation{Institute of Physics, Johannes Gutenberg University of Mainz}
\affiliation{Helmholtz Institute Mainz, 55099 Mainz, Germany}
\affiliation{GSI Helmholtzzentrum für Schwerionenforschung, 64291 Darmstadt, Germany}
\author{Chengtong Zhang}
\affiliation{Department of Chemistry, New York University, New York, NY 10003, United States}
\author{Gary P. Centers}
\affiliation{Institute of Physics, Johannes Gutenberg University of Mainz}
\affiliation{Helmholtz Institute Mainz, 55099 Mainz, Germany}
\affiliation{GSI Helmholtzzentrum für Schwerionenforschung, 64291 Darmstadt, Germany}
\author{Dmitry Budker}
\affiliation{Institute of Physics, Johannes Gutenberg University of Mainz}
\affiliation{Helmholtz Institute Mainz, 55099 Mainz, Germany}
\affiliation{GSI Helmholtzzentrum für Schwerionenforschung, 64291 Darmstadt, Germany}
\affiliation{Department of Physics, University of California, Berkeley, CA 94720, USA}
\author{Danila A. Barskiy}
\altaffiliation{Current address: Frost Institute for Chemistry and Molecular Science, 
Department of Chemistry, University of Miami, Coral Gables, FL 33146, USA}
\affiliation{Institute of Physics, Johannes Gutenberg University of Mainz}
\affiliation{Helmholtz Institute Mainz, 55099 Mainz, Germany}
\affiliation{GSI Helmholtzzentrum für Schwerionenforschung, 64291 Darmstadt, Germany}
\author{Alexej Jerschow}
\email{alexej.jerschow@nyu.edu}
\affiliation{Department of Chemistry, New York University, New York, NY 10003, United States}

\date{\today}

\maketitle
\setstretch{1.5}
\clearpage


This Supporting Information contains additional experimental data and description of data-analysis protocols referenced in the main text (all notebooks are provided). The document is organized as follows: 

\begin{itemize}
    \item Section I: Numerical simulations of the time-dependent magnetic fields experienced by the spin system during shuttling. 
    \item Section II: Simulated nutation profiles following adiabatic or sudden transfer of the [$^{13}$C] formic-acid spin system.
    \item Section III: Global fit of the [$^{13}$C] formic-acid nutation-decay data with the theoretical model described in the main text.
    \item Section IV: Measured $R_{2}^*$ values for water and [$^{13}$C] formic-acid experiments.
    \item Section V: Model extension to the heteronuclear four-spin system of the methyl group in [$^{13}$C] methanol.
    \item Section VI: Nutation results from a different experimental setup than used in the main text. Adiabatic vs. sudden state preparation is briefly investigated.
\end{itemize}

\newpage
\section{Simulating polarization transfer during shuttling}\label{sec:SI_shuttle}

As discussed in the main text, knowledge of the density operator after shuttling is a first key element in understanding the evolution of polarization during the storage time followed by pulse-acquire. Thus, we developed a Spinach simulation that propagates spin systems from the prepolarizing field in the Halbach array to the measurement field inside the mu-metal shield (see notebook ZULF\_BELT\_shuttling\_profile.m). In order to map the spatial dependence of the magnetic field along the shuttling path, we first used a gaussmeter and a fluxgate magnetometer to measure the transverse and longitudinal components, respectively, with zero current applied to the solenoid. Then, we assumed a constant shuttling velocity that allows travel of the path from the Halbach array to the measurement field in a shuttling time of 0.1\,s and generated the time-dependent field profiles (Fig.~\ref{fig:FigS1}a), which we fitted with three independent sigmoid functions (orange lines). With this information, we constructed a total Hamiltonian containing time-dependent Zeeman contributions, a constant Zeeman term along the shuttling path generated by the solenoid field (current through the solenoid tuned to generate the desired constant field $B_0$), and a constant scalar contribution ($J_{\text{CH}}=222$\,Hz) for the [$^{13}$C] formic-acid system containing two hetereonuclear {$^1$H-$^{13}$C} spins:

\begin{equation}
\begin{aligned}
\hat{H}_{\text{tot}}^{\text{shuttling}}\left(t\right) 
&= \hat{H}_{\text{Zeeman}}^{\text{shuttling}}\left(t\right) 
  + \hat{H}_{\text{Zeeman}}^{\text{solenoid}}
  + \hat{H}_{J_{\text{CH}}} \\
&= -\sum_{\mu=x,y,z} B_{\mu}\left(t\right)
   (\gamma_{\text{C}} \hat{I}_\mu + \gamma_{\text{H}} \hat{S}_{\mu})
   - B_0(\gamma_{\text{C}} \hat{I}_z + \gamma_{\text{H}} \hat{S}_z)
   + 2\pi J_{\text{CH}}\,\hat{\mathbf{I}}\cdot\hat{\mathbf{S}} \,.
\end{aligned}
\end{equation}
where we have set $\hbar=1$ for convenience.
Next, we propagated the initial equilibrium polarization characteristic to the initial field in the Halbach array and followed three types of descriptors of the evolving density operator (Fig.~\ref{fig:FigS1}c). We notice the conversion of 1-spin order into 2-spin order, the re-orientation of the polarization axis from the Halbach array's field alignment (perpendicular to the shuttling path) to the solenoid's field alignment (parallel to the shuttling path), and the equilibration of spin density touching both spins. Notably, starting from 0.06\,s, the density operator does not evolve under the coherent Hamiltonian, reaching a stable state. To evaluate whether the transport was adiabatic, we compare the final ratio of the 1-order ($\hat{D}_{z}$ \& $\hat{Z}_{z}$) and 2-order ($\hat{Z}_{x}$) contributions with the one predicted by Eq.~13 in the main text. For the final measurement field used, $B_0=0.1\,\upmu$T ($\theta\approx\pi/4$), the predicted ratio is 
\begin{equation}
    r=\frac{\delta-\epsilon \cos(2\theta)}{\epsilon \sin(2\theta)}\approx \frac{\gamma_{\text{H}}+\gamma_{\text{C}}}{\gamma_{\text{H}}-\gamma_{\text{C}}}=1.66 \,,
\end{equation}
in agreement with the simulation-derived ratio of 1.6, confirming that the shuttling was performed adiabatically. 

To further confirm the adiabaticity of transport, we tested whether the final density operator commutes with the Hamiltonian at the measurement field containing constant Zeeman and scalar interactions ($\hat{H}_{\text{Zeeman}}^{\text{solenoid}}+\hat{\text{H}}_{J_{\text{CH}}}$). We propagated the density operator under this Hamiltonian, projected its evolution onto $\hat{\rho}_{z}=\gamma_{\text{C}}\hat{I}_{z}+\gamma_{\text{H}}\hat{S}_{z}$, and ran a Fourier transformation (this arrangement mimics an experiment with a sensor measuring along the shuttling path). Fig.~\ref{fig:FigS1}b shows no $J$ peaks, confirming that the density operator after shuttling commutes with the total Hamiltonian at the measurement field and, consequently, that the polarization transport was indeed adiabatic. 

\renewcommand{\thefigure}{S\arabic{figure}}
\begin{figure}[h]
\includegraphics[trim=0pt 0pt 0pt 0pt, clip, width=\textwidth]{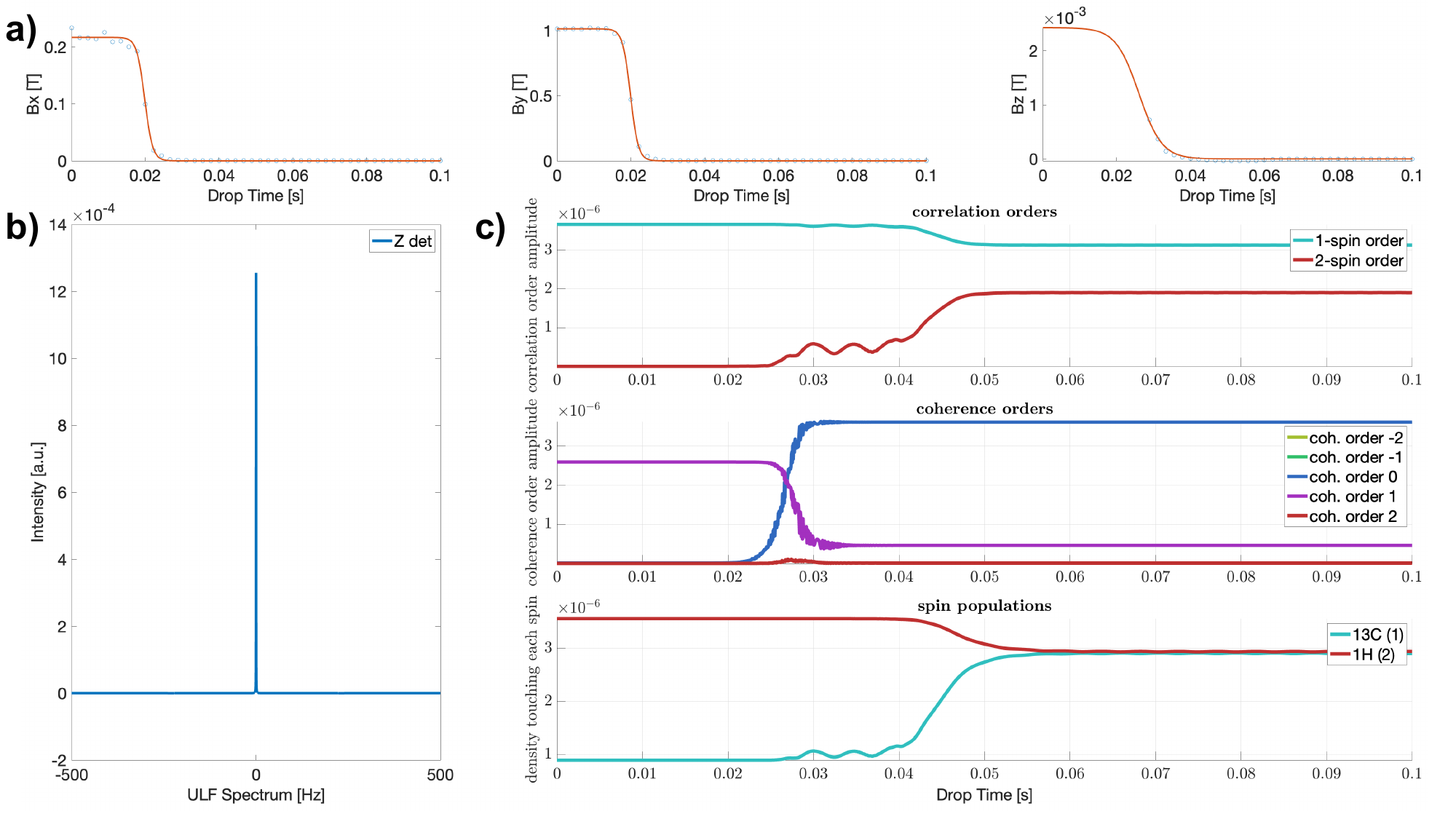}
\caption{
a) Three-dimensional components of the time-dependent magnetic fields experienced by the spin system during shuttling. The orange lines represent fitted sigmoid functions on the experimental points. b) Simulated spectrum immediately after shuttling by allowing the final density operator to evolve under the measurement Hamiltonian ($\hat{H}_{\text{Zeeman}}^{\text{solenoid}}+\hat{H}_{J_{\text{CH}}}$) and projecting its trajectory onto $\hat{\rho}_{z}=\gamma_{\text{C}}\hat{I}_{z}+\gamma_{\text{H}}\hat{S}_{z}$. The absence of $J$ peaks indicates that the transfer was performed adiabatically. c) Density-operator analysis of polarization evolution during  shuttling. 
}
\label{fig:FigS1}
\end{figure}

\section{Simulated nutation profiles for adiabatic and sudden transfer}\label{sec:nut_ad_sud}
\begin{figure}[h]
\includegraphics[trim= 340pt 0pt 0pt 0pt, clip, width=0.8\textwidth]{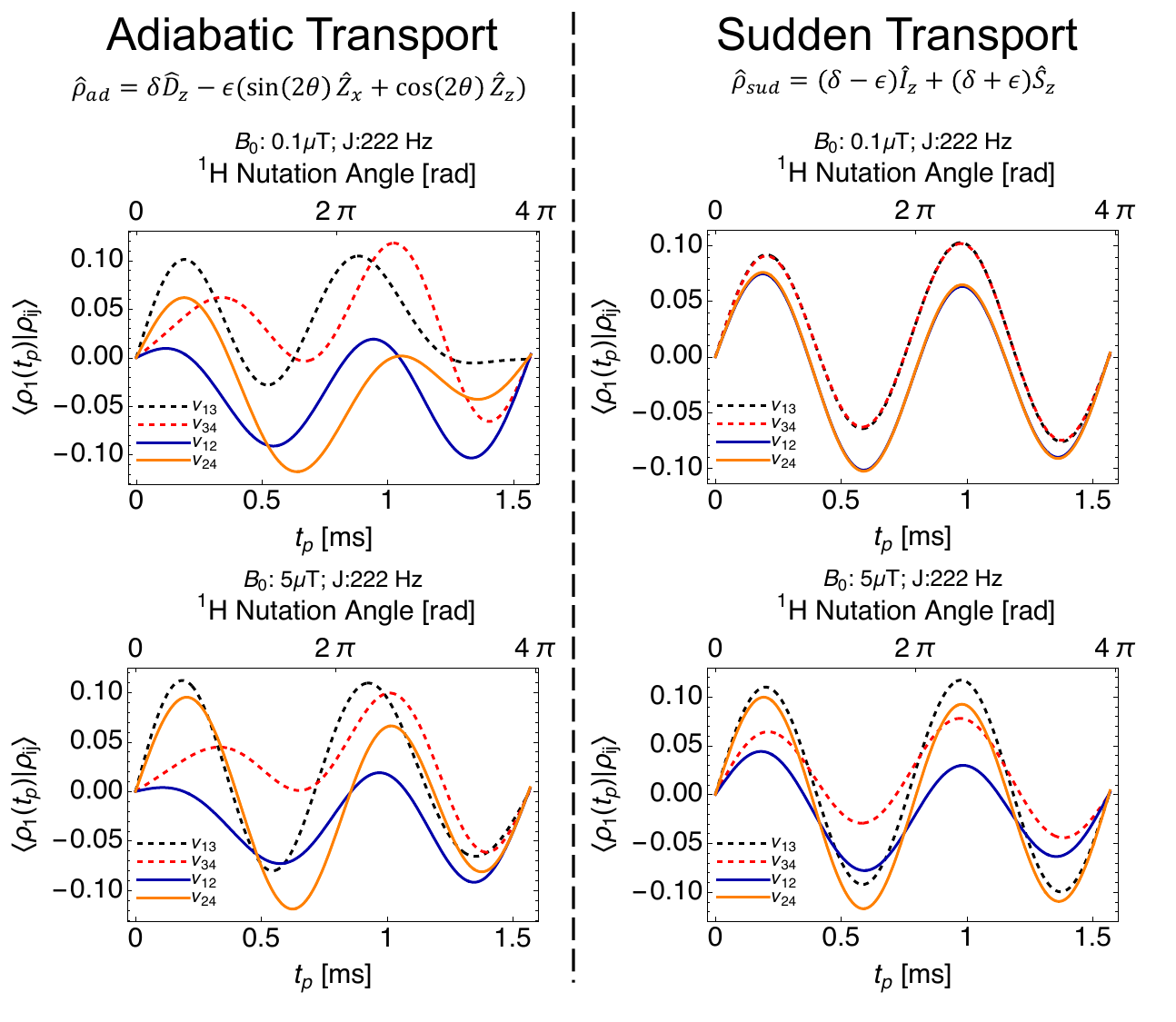}
\caption{Simulated nutation profiles of the four detectable transitions $\rho_{ij}=\ket{\psi_i}\bra{\psi_j}$ characteristic to the $^{13}$C-$^1$H spin system of [$^{13}$C]-formic acid, at two measurement fields (0.1 $\upmu$T and 5$\upmu$T), 
starting from two initial density operators specific to either adiabatic ($\hat{\rho}_{\text{ad}}$) or sudden ($\hat{\rho}_{\text{sud}}$) transport during shuttling. Calculations were performed identically to the ones in Fig.~3d in the main text. 
}
\label{fig:FigS2}
\end{figure}

Another test for evaluating population transfer during shuttling is by analyzing the nutation profiles of the observable peaks. As depicted in Fig.~\ref{fig:FigS2}, these profiles show significant differences for the case of adiabatic and sudden transfer. Moreover, these profiles change with the solenoid field, which imposes different eigenstates. The experimental nutation curves (Figs.~4 \& 5 in the main text), resemble more the case of adiabatic transport, supporting our assumption that populations have been preserved in the instantenous eigenstates during shuttling. 

\newpage
\section{Data Processing and Global Fits of Nutation-Decay Data}\label{sec:global_fit}

The time-domain signals from the two zero-field magnetometers (QuSpin QZFM Gen-2) were processed as in \cite{fabricant_proton_2024}, with spectra collected as the average of 64 scans. Peak ranges were selected for calculating amplitude by including the entire peak shape and excluding noise regions.
This analysis used a zero-filling factor of $2^{17}$ to obtain higher-resolution spectra. We also applied a phase optimization based on minimizing the root mean square error (RMSE) of the Lorentzian-function fit residuals. For spectral analysis, an expected resonance-frequency range was selected for each peak of interest. Baseline correction was applied by subtracting a linear fit to the average of each of two 1\,Hz regions on either side of the resonance peak. From the fitting parameters, we extracted the peak amplitude and the full width at half maximum (FWHM) with error bars propagated from the fitting uncertainty ($1\sigma$ confidence interval) of each parameter. The reported peak-amplitude uncertainty captures the run-to-run statistical and systematic variation but does not incorporate other systematics such as sensor-gain and sample-geometry variability, which were determined to be of negligible importance for the results reported in this work.

\begin{figure}[h]
\includegraphics[trim=0pt 70pt 0pt 1pt, clip, width=\textwidth]{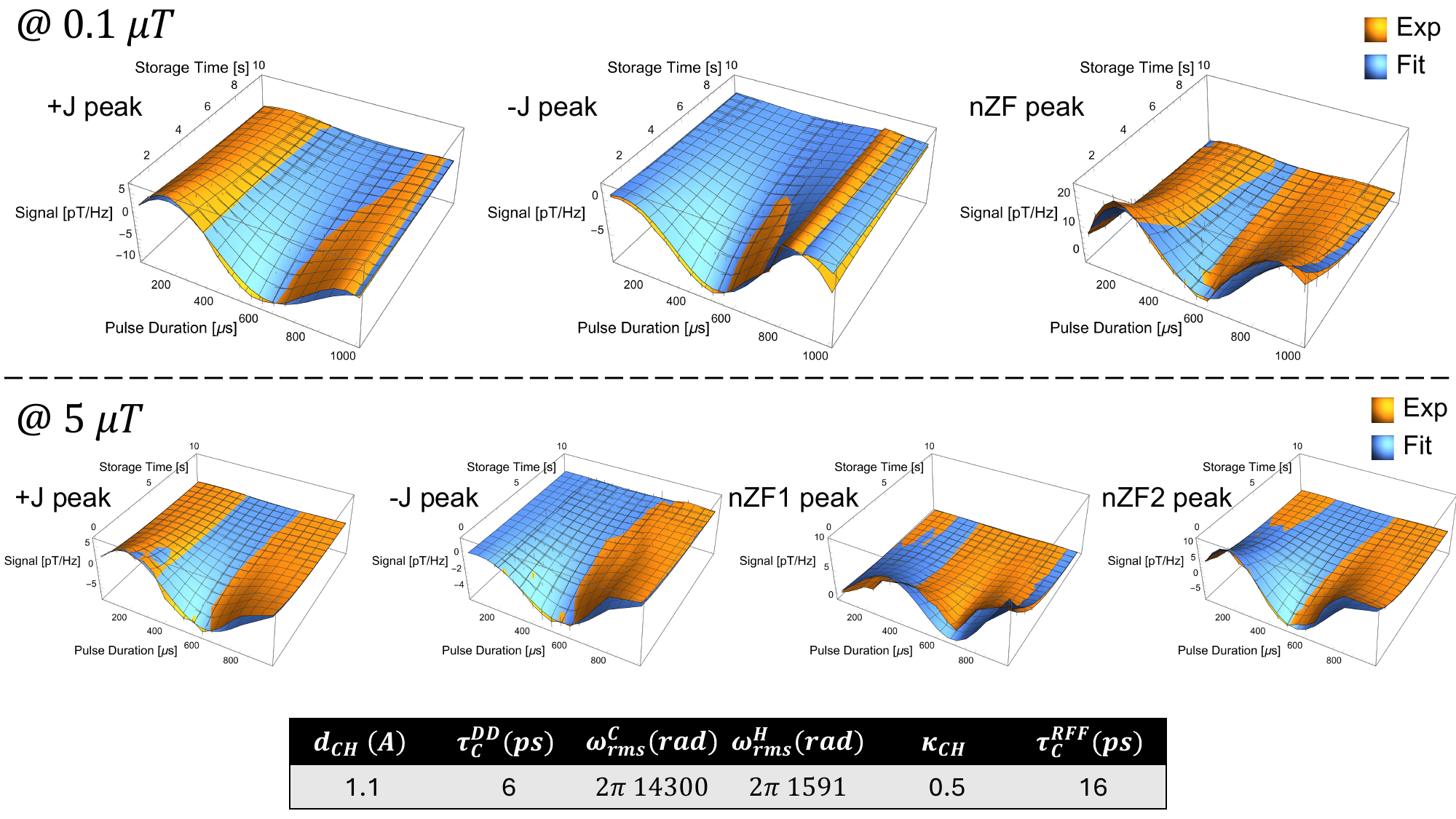}
\caption{Overlay of the experimental peak amplitudes at different nutation pulse lengths $t_p$ and storage times $t_s$ (orange) at both fields and the best fitted model (blue) with five fitting variables summarized in Table\,1. 
}
\label{fig:FigS3}
\end{figure}

The experimental pulse sequences shown in Fig.~4c allowed us to measure the signal amplitude of each detectable peak at two measurement fields ($0.1\,\upmu$T and $5\,\upmu$T) as a function of both storage time $t_s$ and detection pulse length $t_p$, leading to a dataset of seven three-dimensional surfaces (Fig.~\ref{fig:FigS3}). At $0.1\,\upmu$T, we monitored the $\pm J$ peaks ($\nu_{12}$ and $\nu_{24}$ transitions) and the sum of the two nZF peaks ($\nu_{13}$ and $\nu_{34}$ transitions), as the separation of the latter at this field was smaller than the characteristic linewidths. For a $5\,\upmu$T field, we were able to track all four peaks individually. 

For each measurement field, we fitted the experimental nutation-decay surfaces simultaneously using the attached script (R$_1$\_ULF\_global\_fit.nb).
We started by fitting the nutation curves for zero storage time ($t_s = 0$\,s) in order to extract the maximum amplitude of the signal for each decay curve of individually monitored peaks at different nutation pulse lengths. To do so, we extended Eqs.\,3--6
in the main text to incorporate also the evolution due to the scalar coupling during the nutation pulse, as it becomes relevant for long pulses ($t_p \geq 1/4J_{CH}$).
The total Hamiltonian acting on the initial density operator $\hat{\rho}_0(t_s)=\sum_{i=1,4}{p_{\psi_{i}}(t_s)\ket{\psi_{i}}\bra{\psi_{i}}}$ becomes
\begin{equation}
\hat{H}_{\text{tot}}^{\text{pulse}}=-B_{y}^{\text{pulse}}(\gamma_\text{C} \hat{I}_y +\gamma_\text{H} \hat{S}_y) + 2\pi J_{\text{CH}}\hat{\textbf{I}}\cdot\hat{\textbf{S}} \,.
\end{equation}
Due to the complexity of the resulting Hamiltonian, analytical expressions cannot be derived. Therefore, we substituted the known parameters such as the magnetic field of the applied pulse $B_{p}$, the solenoid field $B_0$, and the scalar coupling constant $J_{\text{CH}}$, and numerically propagated the density operator $\hat{\rho}_0$ using the sandwich equation $e^{-i\hat{H}_{\text{tot}}^{\text{pulse}}t_p}\hat{\rho}_0\left(t_s\right)e^{i\hat{H}_{\text{tot}}^{\text{pulse}}t_p}$. We computed individual nutation curves by projecting the resulting operator on the transition operators $\ket{\psi_{i}}\bra{\psi_{j}}$. Thus, we numerically derived accurate equations to fit the measured nutation profiles which depend on the resulting populations $p_{\psi_{i}}(t_s)$ following mixing during storage time. 

To model population evolution after shuttling, we considered the two sets of coupled differential equations (CDFs) characteristic to field-dependent relaxation mechanisms given by the $^{13}$C-$^1$H dipolar interaction (DD) and random field fluctuations (RFF) at the two nuclei (see attached notebook R$_1$\_ULF\_global\_fit.nb). These equations include both auto- and cross-relaxation but no cross-correlation between the two mechanisms. The total evolution of states' populations is given by the additive sum of the two mechanisms and is dependent on six molecular parameters: the $^{13}$C-$^1$H distance ($d_{\text{CH}})$, the rotational correlation time of the dipolar interaction ($\tau_{c}^{\text{DD}}$), the amplitudes of the random field fluctuation at the two nuclei ($\omega_{\text{rms}}^{\text{H}}$ and $\omega_{\text{rms}}^{\text{C}}$), a constant describing the correlation between the two fluctuations ($\kappa_{\text{CH}}$), and the correlation time of the RFFs ($\tau_{c}^{\text{RFF}}$). Our goal was to estimate these parameters by fitting the experimental data with our model. However, the system of CDFs cannot be solved analytically, so we implemented an iterative least-square fitting algorithm to extract these quantities (see attached notebook R$_1$\_ULF\_global\_fit.nb).
Figure~\ref{fig:FigS3} shows the overlay of the experimental data (orange) and the best fit model (blue) after simultaneously fitting all available data with a single model. The best-fit estimates of the unknown molecular quantities leading to relaxation are summarized in Table\,1 in the main text.

\section{Transverse Relaxation Rates $\boldsymbol{R_2}$ at 0.1 $\boldsymbol{\upmu}$T}\label{sec:R2_water}

In order to derive an approximate experimental value of the transverse relaxation rates,  
we fitted the Lorentzian lineshape of individual peaks and extracted the FWHM. This, in turn, is related to the transverse relaxation rate of the corresponding transition by $R_{2}^{*}=\pi\cdot\text{FWHM}$. 
Figure~\ref{fig:FigS3}a shows the experimentally derived $R_2^{*}$ values for the proton peak of a water sample as a function of external field. As no relaxation dispersion is expected---the eigenstates of the two-spin-1/2 system do not change with increasing field because the two protons are always chemically and magnetically equivalent---the increase in apparent transverse rate is attributed to higher field inhomogeneities present in our setup. Given evidence for field inhomogeneities due to solenoid leakage and gradients from the zero-field magnetometer compensation coils, we expect that relaxation-rate values derived from peak linewidths should be significantly larger than the intrinsic transverse relaxation rates imposed by spin interactions, especially at higher fields.

\begin{figure}[h]
\includegraphics[trim=0pt 100pt 0pt 100pt, clip, width=\textwidth]{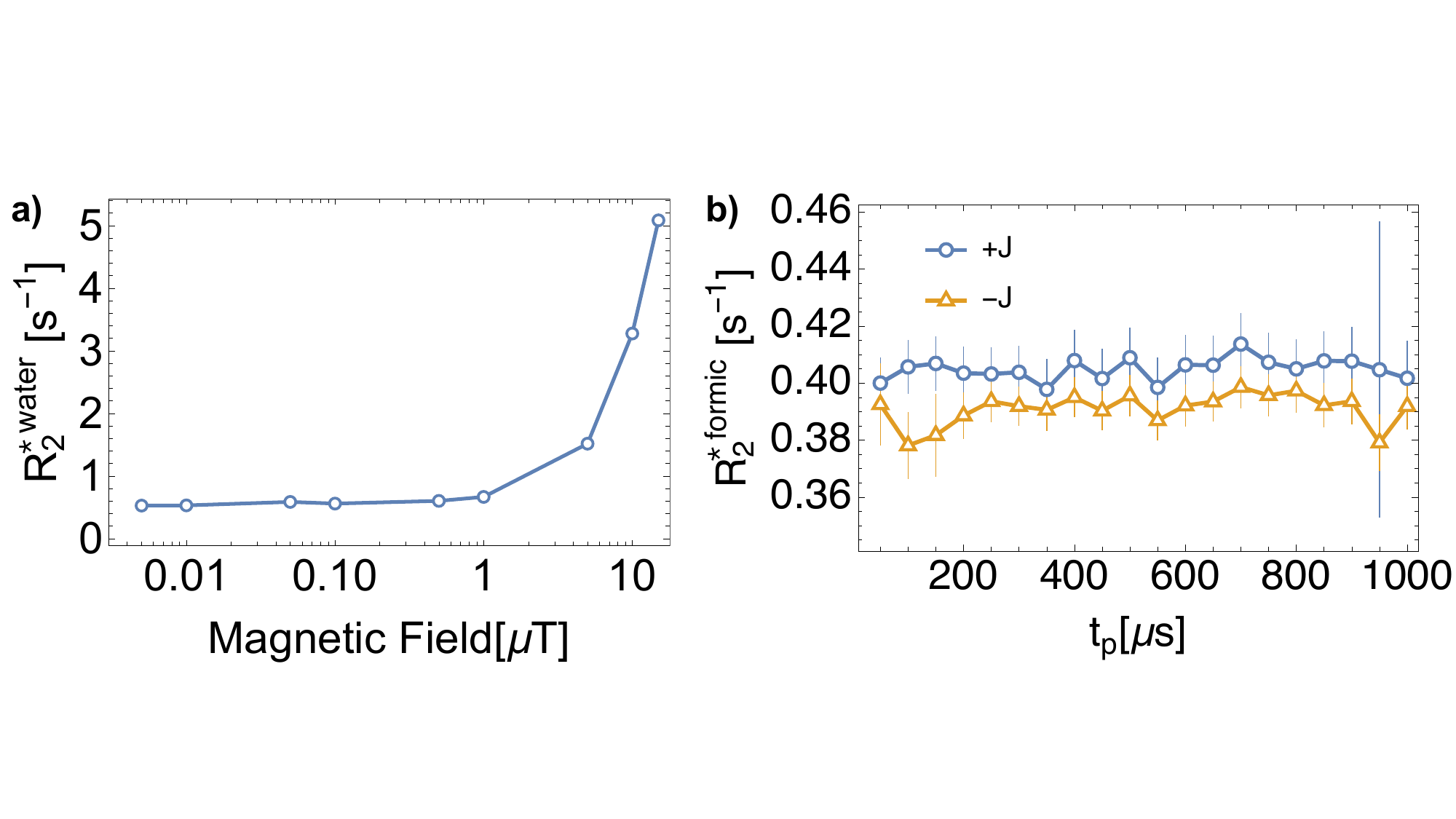}
\caption{a) Field dependence of apparent transverse relaxation rate $R_2^*$ 
of the proton peak in a water sample. The changes are attributed to increasing field inhomogeneities. Error bars are smaller than the size of data points.
b) Transverse relaxation rates $R_2^*$ for different $[^{13}\mathrm{C}]$ formic-acid resonance peaks at $0.1\,\upmu\mathrm{T}$ with $1$\,s storage time. 
Higher uncertainty at $950\,\upmu\mathrm{s}$ corresponds to lower-intensity peaks, as seen in the nutation curve (Fig.~4d).}
\label{fig:FigS4}
\end{figure}

Figure~\ref{fig:FigS4}b shows the $R_2^*$ rates 
of the $\pm J$ peaks for the [$^{13}$C] formic-acid sample measured at $0.1\,\upmu$T using different pulse lengths. We highlight the relatively constant behavior of $R_2^*$ as opposed to the apparent decay rates during storage time ($R_1$), which were shown to be dependent on the induced nutation angle (see main text). 

\section{Extending the theoretical model to $^{13}$C-methanol spin systems}
\label{sec:13CH3}
\begin{figure}[h]
\includegraphics[trim=0pt 0pt 0pt 00pt, clip, width=\textwidth]{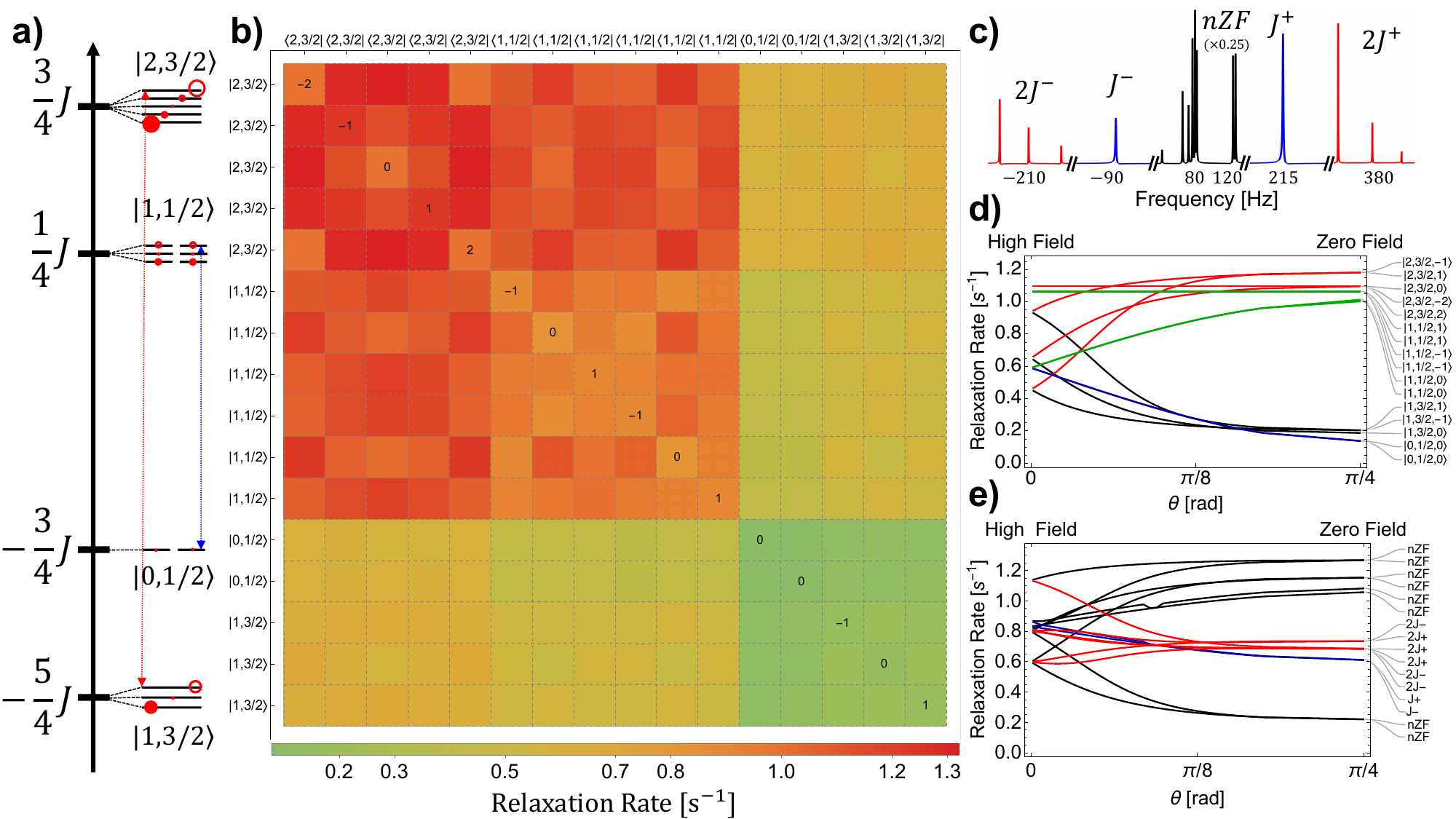}
\caption{a) Energy-level diagram for a [$^{13}$CH$_3$] system at ultralow field ($B_{0}=2.3\,\upmu$T and $J_{\text{CH}}=140$\,Hz). States' populations are schematically represented (overpopulated with filled 
and underpopulated with empty 
red circles) assuming an adiabatic transfer of high-field polarization. Observed transitions connect spin states satisfying the selection rule $\Delta m_{f} = \pm 1$ and are labeled as follows: nZF-peaks connect states inside the same manifold, $J$-peaks connect $\ket{0,1/2}$ and $\ket{1,1/2}$ manifolds, and $2J$-peaks connect $\ket{2,3/2}$ and $\ket{1,3/2}$ manifolds. b) Matrix plot of auto-relaxation rates of all populations (diagonal terms) and coherences (off-diagonal terms) among the [$^{13}$CH$_3$] system's eigenstates at zero field. Axes labels encode the values of the \textit{F} and \textit{S} quantum numbers and the insets on the diagonal encode $m_{F}$ values. Eigenstates are ordered by increasing energy (from right to left). Only dipolar interactions were considered. c) Example of simulated ULF spectrum of [$^{13}$CH$_3$] using quadrature detection ($B_{0}=2.3\,\upmu$T and $J_{\text{CH}}=140$\,Hz). The maximum spectral complexity is achieved with $2^{4}$ observable peaks. (d,e) Field dependence of the auto-relaxation rates of the eigenstates' populations (d) and observable coherence (e). Simulation parameters: $\tau_{\text{C}}^{\text{DD}}=10$\,ps, $r_{\text{CH}}=1.1$\,\AA, $\theta_{HCH}=70.5\,^{\circ}$.}
\label{13CH3}
\end{figure}

Analytical modeling becomes impractical for larger spin systems, such as [$^{13}$CH$_3$], but the workflow can be easily scaled and the equations describing spin dynamics can be solved numerically. Figure~\ref{13CH3} summarizes the most important aspects characterizing relaxation phenomena for the case of the $^{13}$CH$_3$ system where dipole-dipole interactions dominate. First, the coherent Hamiltonian describing the system is: 
\begin{equation}
        \hat{H}_{0} = -\left(\gamma_{\text{C}}\hat{I}_{z}+\sum_{i=1,2,3}{\gamma_{\text{H}}\hat{S}_{i,z}}\right)B_{0} + 2\pi J_{IS} \sum_{i=1,2,3}\boldsymbol{\hat{I}}\cdot\boldsymbol{\hat{S_i}} \,.
\end{equation}
Due to the strong coupling of heteronuclei given the ultralow field, we label the states according to three quantum numbers as $\ket{F,S,m_{f}}$, indexing the total spin angular momentum (\textit{F}), the spin angular momentum of the $^{1}$H system (\textit{S}), and the projection of the total spin angular momentum ($m_{F} =-F, -F + 1, \ldots, F$)\cite{stern_simulation_2023}. The energy-level diagram is shown in Fig.~\ref{13CH3}a, with states populated assuming an adiabatic transport of high-field polarization of the [$^{13}$CH$_3$] system. Similar to [$^{13}$C-$^{1}$H] described above, a complex exchange and decay of populations takes place during storage time which can be modeled numerically. Figure~\ref{13CH3}b indicates the auto-relaxation rates of all populations (diagonal elements) and coherences (off-diagonal elements) that are created by the eigenstates of the spin system in zero field. Two matrix blocks are apparent: one defined by the low-energy $\ket{1,3/2}$ and $\ket{0,1/2}$ manifolds with lower relaxation rates, and the other by high-energy $\ket{1,1/2}$ and $\ket{2,3/2}$ manifolds with higher relaxation rates. This clear segregation in terms of polarization lifetimes is specific to ultralow fields, while higher fields tend to correspond to more evenly distributed relaxation rates (Fig.~\ref{13CH3}d). A similar trend is observed for the lifetime of observable polarization during acquisition, with ultralow-field peaks having significantly different linewidths which converge at high fields (Fig.~\ref{13CH3}e). Another particularity of ULF spectra is the complexity in terms of detectable non-degenerate transitions, with a total of 16 individual resonances as compared to only six ($^{13}$C quartet and $^1$H doublet) at high fields\cite{appelt_paths_2010}. This allows for a better estimation of parameters describing the relaxation mechanisms at play, given that more observables can be used to fit a specific model. 

\section{Nutation-curve reproducibility and state preparation}\label{sec:sudden_transfer}

In this section we reproduce the nutation curve demonstrated in the main text (see Fig.~3d) using a different ZULF-NMR experimental setup and briefly investigate various pulse/detection geometries and state preparation. 

\begin{figure}[h!]
	\includegraphics[width=\textwidth]{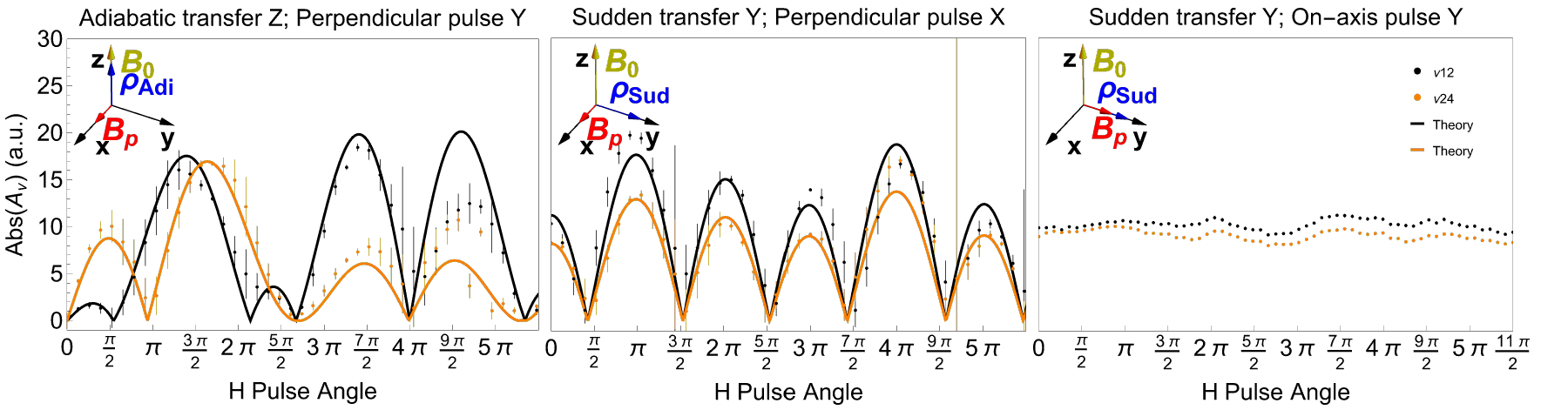}
	\caption{Comparison of nutation theory for various state-preparation and pulse-axis geometries. The detection is single-channel and always oriented along the $y$-axis. Left: Adiabatic preparation and perpendicular pulse, reproducing and extending the data shown in Fig.~5d of the main text. Center: Sudden preparation and perpendicular pulse, which shows strong agreement with the theory. Right: Sudden preparation and on-axis pulse, demonstrating a ``null'' result limited by pulse/detector alignment and/or state-preparation quality.}
	\label{fig:gsm1}
\end{figure}

The $[^{13}\mathrm{C}]$ formic-acid sample was thermally polarized at 2\,T for 30\,s then shuttled to the detection field of $B_0\approx54$\,nT. The field-transfer profiles to this detection field were controlled to provide either adiabatic or sudden state preparation along a chosen axis. The pulse amplitude was $B_p=30.64\,\upmu$T and the pulse duration was varied as $t_p=[0,2100]\,\upmu$s in steps of $50\,\upmu$\,s (for the on-axis pulse, $B_p=20.8\,\upmu$T and $t_p=[0,3200]\,\upmu$s). These pulse amplitude and duration sweeps correspond to a proton pulse angle, $\theta_\text{H}$, from 0 to $\approx11\pi/2$. Detection was performed by one channel of a QuSpin magnetometer (QZFM Gen-2) oriented along the $y$-axis (see inset coordinates of Fig.~\ref{fig:gsm1}; the geometry is analogous to that shown in Fig.~3a of the main text). Each data point in Fig.~\ref{fig:gsm1} is the fitted amplitude of the respective coherence from an average of eight scans of 20\,s acquisition time at each pulse length. The error bars are the parameter standard errors from the fit obtained from the estimated asymptotic covariance matrix of a nonlinear least-squares minimization in the Fourier domain, using Mathematica (the standard NonlinearModelFit[] function).

The left plot of Fig.~\ref{fig:gsm1} demonstrates excellent agreement with Eqs.~3--6 of the main text. The pulse duration was extended to 2200\,$\upmu$s ($\theta_H\approx11\pi/2)$, with deviations from theory after 1600\,$\upmu$s potentially explained by a combination of pulse imperfections, magnetic noise, $J$-coupling evolution during pulse, and/or cross-relaxation.

For sudden transfer, the sample is adiabatically oriented 
in the detection region to a holding field of $\approx30\,\upmu$T along a chosen axis. This field is suddenly turned off in less than $10\,\upmu$s), preparing a density matrix proportional to that of Eq.~3 of the main text, $\hat\rho_{\text{sud}}\propto(\delta+\epsilon)\hat I_j+(\delta-\epsilon)\hat S_j$, with $j$ being the chosen axis of the holding field. This density matrix can be used to derive the nutation curves as outlined in the main text, giving
\begin{widetext}
	\begin{align}
		A_{\nu_{12},\nu_{24}}(t_p) \propto (\epsilon-\delta)\cos{(B_{p}\gamma_{\text{C}}t_{p})} + (\epsilon+\delta)\cos{( B_{p}\gamma_{\text{H}}t_{p})}.
	\end{align}
\end{widetext}
The center plot of Fig.~\ref{fig:gsm1} demonstrates excellent agreement with this nutation equation, with the initial experimental amplitudes of the $\pm J$ peaks ($\nu_{24}$ or $\nu_{12}$) determining the scaling and $\delta,\epsilon$ set to their theoretical values.

Since the pulse Hamiltonian, $\hat H_p=(\gamma_I\hat I_k+\gamma_S\hat S_k)B_p$, has the same form as the sudden-transfer density matrix, one expects no evolution due to the pulse Hamiltonian if the preparation and pulse axis are the same ($j=k$). This ``null'' measurement is shown in the right plot of Fig.~\ref{fig:gsm1}. Performing this experiment allows one to characterize the quality of their state preparation and/or pulse-coil/detection axes alignments. If the residual variation is correlated between the $\pm J$ coherences (as in the central plot), this indicates a misalignment between the coil/detection axes. The residual variation can be used to estimate the projection of the pulse along the perpendicular axis. If the residual variation is not correlated (and perhaps follows a trend described by an adiabatic preparation and on-axis pulse nutation) this indicates a mixture of sudden and adiabatic state preparation. This could be caused, for example, by a too-low slew rate of the pulse-coil voltage or a too-long time constant of the RL circuit. The experimental results from this particular setup have correlated residuals, indicating an alignment error of $\lesssim 5^\circ$.

\subsection{State preparation}
As discussed in the previous section, the density-matrix preparation before the pulse is determined by either adiabatic or sudden transfer to the detection field. Figure~S7 shows the results of a measurement where a holding field on-axis to the detection was linearly dropped to zero field at different rates. For sudden transfer, no pulse is required as the Zeeman eigenstates evolve under the $J$-coupling Hamiltonian. In contrast, adiabatic preparation requires a pulse for observable magnetization. The sudden-transfer signal amplitude as a function of field drop rate allows one to determine the required threshold for sudden preparation of states.

\begin{figure}[h!]
    \includegraphics[scale=1]{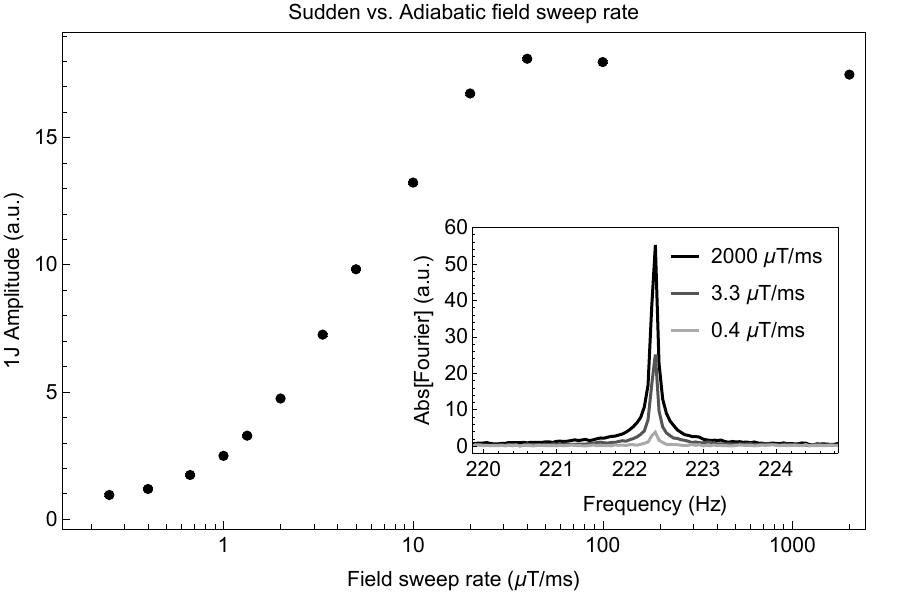}
	\caption{Measurement data of thermally polarized (at $\approx2$\,T) [$^{13}$CH$_3$] at a detection field of zero, where the rate of a linear drop from 20\,$\upmu$T to 0\,$\upmu$T was varied. The amplitude of the $1J$ frequency coherence demonstrates the transition from sudden to adiabatic preparation of states.}
    \label{fig:FigS5}
\end{figure}

\clearpage
\bibliography{references}